\documentclass[twocolumn,pre,aps,showpacs,superscriptaddress]{revtex4-2}
 \usepackage{graphicx}
\usepackage[latin1]{inputenc}
\usepackage[T1]{fontenc}
\usepackage{bbm}
\usepackage{soul}
\usepackage{bm}
\usepackage[usenames]{color}
\usepackage{multirow}
\usepackage{amssymb}
\usepackage{amsbsy}
\usepackage{mathtools}
\usepackage{amsmath}
\usepackage{stmaryrd}
\usepackage{graphicx}
\usepackage{epsfig}
\usepackage{placeins}
\usepackage{bbold}
\usepackage{mathrsfs}
\usepackage{braket}
\usepackage{blindtext}
\usepackage[colorlinks,linkcolor=blue,citecolor=blue,urlcolor=blue]{hyperref}
\usepackage{scalerel}
\usepackage{tikz}
\usetikzlibrary{calc}
\usetikzlibrary{patterns}
\usetikzlibrary{svg.path}
\definecolor{orcidlogocol}{HTML}{A6CE39}
\tikzset{
  orcidlogo/.pic={
    \fill[orcidlogocol] 
svg{M256,128c0,70.7-57.3,128-128,128C57.3,256,0,198.7,0,128C0,57.3,57.3,0,128,
0C198.7,0,256,57.3,256,128z};
    \fill[white] svg{M86.3,186.2H70.9V79.1h15.4v48.4V186.2z}
                 
svg{M108.9,79.1h41.6c39.6,0,57,28.3,57,53.6c0,27.5-21.5,53.6-56.8,
53.6h-41.8V79.1z 
M124.3,172.4h24.5c34.9,0,42.9-26.5,
42.9-39.7c0-21.5-13.7-39.7-43.7-39.7h-23.7V172.4z}
                 
svg{M88.7,56.8c0,5.5-4.5,10.1-10.1,10.1c-5.6,0-10.1-4.6-10.1-10.1c0-5.6,4.5-10.1
,10.1-10.1C84.2,46.7,88.7,51.3,88.7,56.8z};
  }
}

\newcommand\orcid[1]{\!%
  \href{https://orcid.org/#1}{%
    \mbox{%
      \scaleto{%
        \begin{tikzpicture}[yscale=-1,transform shape]
          \pic{orcidlogo};
        \end{tikzpicture}
      }{8pt}%
    }%
  }%
}

\newcommand{\jz}[1]{{\color{black} #1}}
\begin{document}

\title{Eigenstate Thermalization Hypothesis and Random Matrix Theory Universality in Few-Body Systems}

\author{Jiaozi Wang~\orcid{0000-0001-6308-1950}}
\email{jiaowang@uos.de}
\affiliation{Department of Mathematics/Computer Science/Physics, University of Osnabr\"uck, D-49076 
Osnabr\"uck, Germany}

\author{Hua Yan~\orcid{0000-0002-3199-4648}}
\email{yanhua@ustc.edu.cn}
\affiliation{CAMTP - Center for Applied Mathematics and Theoretical Physics,
University of Maribor, Mladinska 3, SI-2000 Maribor, Slovenia}
\author{Robin Steinigeweg~\orcid{0000-0003-0608-0884}}
\affiliation{Department of Mathematics/Computer Science/Physics, University of Osnabr\"uck, D-49076 
Osnabr\"uck, Germany}

\author{Jochen Gemmer~\orcid{0000-0002-4264-8548}}
\affiliation{Department of Mathematics/Computer Science/Physics, University of Osnabr\"uck, D-49076 
Osnabr\"uck, Germany}

\date{\today}

\begin{abstract}
In this paper, we use the Feingold--Peres model, a well-known paradigm of quantum chaos, as an example. Using semiclassical analysis and numerical simulations, we investigate the statistical properties of observables in few-body systems with chaotic classical limits and the emergence of random matrix theory universality.
More specifically, we focus on: (1) the applicability of the eigenstate thermalization hypothesis in few-body systems and the dependence of its form on the effective Planck constant and (2) the existence of a universal random matrix theory description of observables when truncated to a small microcanonical energy window.
Our results provide new insights into the established field of few-body quantum chaos and help bridge it to modern perspectives, such as the general eigenstate thermalization hypothesis (ETH).

\end{abstract}

\maketitle



\section{Introduction}
One of the most fundamental features of quantum chaotic systems is their similarity to random matrices in terms of fluctuation properties \cite{Haake-chaos-book,Casati-chaos80}. For the spectrum of a system with a chaotic classical limit, it is conjectured that the spectral fluctuations are universal and can be described by the random matrix theory (RMT).
{The conjecture has been analytically verified using semiclassical theory, with a focus on the spectral form factor \cite{Berry-chaos-81,Casati-chaos80,Berry-chaos-85,Sieber-chaos-01,Haake-chaos-book,Kaufman-chaos-79,Mueller-chaos-04,Mueller-chaos-05,Wigner-chaos}. }More recently, the connection between spectral fluctuations and RMT has also been established in quantum chaotic systems without a classical limit \cite{PhysRevX.8.021062_RMT_Prosen,PhysRevLett.121.264101_RMT_Prosen,PhysRevX.8.041019-chaos-sff}.

Unlike spectral statistics, the statistical properties of eigenstates are basis dependent. 
{One of the most commonly considered bases is the configuration basis. In this basis, Berry conjectured \cite{Berry77} that the components of chaotic eigenfunctions follow a Gaussian distribution, consistent with RMT. Berry's conjecture has been further supported by numerical results for quantum billiards \cite{PhysRevA.37.3067-ef-billiard-Kaufman}. 
In studies of the transition between quantum chaos and integrability, another particularly important basis is the unperturbed basis, namely, the eigenbasis of the integrable Hamiltonian. Studies of eigenfunction statistics in the unperturbed basis \cite{Buch-EF-82,Benet-EF-03,Benet-EF-07,Meredith-EF-88,PhysRevE.107.064119,Klaus25} have shown that, after appropriate rescaling by their average shape, the components of chaotic eigenfunctions also exhibit a Gaussian distribution \cite{JZ-EF-18}.}

In addition to the spectrum and eigenstates, statistical properties of matrix elements of observables in the eigenbasis of chaotic systems have also attracted significant attention, especially after the introduction of Eigenstate Thermalization Hypothesis \cite{Srednicki94, Deutsch,rigol2008thermalization}  (ETH).
Considering an observable ${\cal O}$, the ETH ansatz  \cite{Srednicki_1996,Srednicki_1999} 
{postulates a particular structure for its matrix elements}
in the eigenbasis of a generic Hamiltonian ${H}$, 
	\begin{equation}\label{eq-ETH}
	{\cal O}_{\alpha \beta} = O(\overline{E})\delta_{\alpha\beta} + 
	\rho^{-1/2}(\overline{E})f(\overline{E},\omega)r_{\alpha\beta}\  ,
	\end{equation}
	where $\omega = E_\alpha-E_\beta$, $\overline{E} = (E_\alpha + E_\beta)/2$, and ${\cal O}_{\alpha\beta} = 
	\bra{\alpha}{\cal O}\ket{\beta}$ . Moreover, 
	$\rho(\overline{E})$ is the density of states, $O(\overline{E})$ and 
	$f(\overline{E},\omega)$ 
	are smooth functions, 
    $r_{\alpha\beta} = r_{\beta\alpha}^\ast$, whose real and imaginary parts are often assumed to be random variables, with zero mean and unit variance. While ETH remains an assumption, its general features have been numerically confirmed \cite{rigol2008thermalization,rs-10,krrg-12,bmh-14,bmh-15,kpsr-13,shp-13,kih-14,skngg-14,ksg-15,mfsr-16}. It is now widely accepted that ETH is valid
    in many-body quantum chaotic systems for few-body observables. 

In contrast to many-body systems, ETH in few-body systems has received much less attention \cite{David22,wang2022semiclassical,Srednicki_1996}. In Srednicki's original paper \cite{Srednicki94}, the argument was based mainly on Berry's conjecture, which is also expected to hold for chaotic few-body systems. From this perspective, it is natural to expect that some form of ETH may also be applicable in such systems. However, a general theoretical framework has not yet been established, and many important questions remain open. In particular, the form of the envelope function, $f(\overline{E}, \omega)$
especially its dependence on the effective Planck constant, remains an open question. This will be our first main focus.

\jz{More recently, studies have extended the conventional ETH framework by considering higher-order correlations among matrix elements of observables \cite{Chan:2018fsp,goold-otoc,Murthy2019,PhysRevE.102.042127,PhysRevLett.128.180601,PhysRevLett.128.190601,iniguez2023microcanonical}.
A generalized form of ETH, often referred to as the full ETH \cite{FoiniPRE,Pappalardi:2023nsj,pappalardi2023microcanonical,GETH-Foini19,Fava:2023pac}, has been introduced, which describes such correlations in terms of smooth functions. 
The full ETH is based on the assumption that the statistical properties of observable matrix elements are invariant under local unitary, or rotational, transformations.
This local unitary invariance motivates a universal random matrix theory description of observables: when restricted to sufficiently small energy windows, the corresponding truncated operators can be regarded as samples from a unitarily invariant ensemble (UIE).}

\jz{Numerical evidence consistent with such a UIE description has been observed for few-body observables in several many-body chaotic systems \cite{wang2023emergence,GETH-Foini19}.
However, to the best of our knowledge, no studies have yet examined few-body systems with a chaotic classical limit.
It remains an open question whether such a UIE description emerges in these systems, and on what energy scale; this will be our second main focus.}
 
In this paper, using numerical simulations and semiclassical analysis, we investigate the questions stated above in the Feingold-Peres model, which is a well-known paradigm of quantum chaos. 
The rest of the paper is organized as follows: In Sec.~\ref{sec-model}, we introduce the Feingold-Peres model and its classical limit. 
Statistical properties of the spectrum are discussed in Sec.~\ref{sec-spectrum}, including both average and fluctuation properties.
Questions related to the conventional ETH are studied in Sec.~\ref{sec-eth-old}. \jz{In Sec.~\ref{sec-eth-new}, we go beyond the conventional ETH and investigate the UIE description of microcanonically truncated operators.}
Finally, conclusions and discussions are given in Sec.~\ref{sec-conclusion}.

\section{The Model}
\label{sec-model}


We begin by considering a coupled angular momentum system, also known as the Feingold-Peres model \cite{feingold1983regular,peres1984new,peres1984stability,feingold1986distribution}, governed by the Hamiltonian  

\begin{equation}  
H= \frac{1+\lambda}{j+1/2}(L_{z}^{1}+L_{z}^{2})+\frac{4(1-\lambda)}{(j+1/2)^2}L_{x}^{1}L_{x}^{2},  
\end{equation}  
where $\lambda \in [0,1]$ is a parameter that controls the relative strength of the two terms. The operators $L_{x,y,z}^{k}$ represent the angular momentum components at site $k = 1,2$, satisfying the commutation relation $[L_{x}^{k},L_{y}^{k^{\prime}}] = iL_{z}^{k} \delta_{kk^{\prime}}$. Furthermore, the total angular momentum is conserved, given by $(\boldsymbol{L}^{k})^{2} = (L_x^{k})^{2} + (L_y^{k})^{2} + (L_z^{k})^{2}$, {with fixed angular momentum magnitude}. We consider $(\boldsymbol{L}^{1})^{2} = (\boldsymbol{L}^{2})^{2} = j(j+1)$ throughout this paper.
The system is naturally described in the basis  $|m_1,m_2\rangle = |j,m_1\rangle \otimes |j,m_2\rangle$, which constitute the common eigenstates of $L^1_z$ and $L^2_z$,  where
{\begin{equation}
L_{z}^{1}|m_1\rangle=m_1|m_1\rangle,\ L_{z}^{2}|m_2\rangle=m_2|m_2\rangle ,
\end{equation}}
for $m_1, m_2 \in \{-j, -j+1, \dots, j\}$. The dimension of the Hilbert space is given by ${\cal D} = (2j+1)^2$. 

As pointed out in Ref. \cite{fan2017quantum}, the system exhibits several symmetries, including two unitary ones: the exchange of angular momenta and a simultaneous rotation of both momenta by an angle \(\pi\) around the \(z\)-axis. Additionally, the system possesses a spectral mirror symmetry, described by the chiral operator \(C = e^{i\alpha} \mathcal{R}_x^1(\pi) \otimes \mathcal{R}_y^2(\pi)\), where \(e^{i\alpha}\) is a phase factor and \(\mathcal{R}^k_{x,y,z}(\theta)\) represents a rotation by angle \(\theta\) around the respective axis, at site $k$. The combination of these symmetries gives rise to nonstandard symmetries within the framework of the Altland-Zirnbauer tenfold classification of quantum systems \cite{altland1997nonstandard}.
In this paper, we focus on the trivial subspaces formed by the basis
\begin{equation}
    |m_{1}, m_{2},+-\rangle=\frac{|m_{1}m_{2}\rangle+|m_{2}m_{1}\rangle}{\sqrt{2}}
\end{equation}
and 
\begin{equation}
    |m_{1}, m_{2},--\rangle=\frac{|m_{1},m_{2}\rangle-|m_{2},m_{1}\rangle}{\sqrt{2}},
\end{equation}
where $m_2>m_1$ and $2j - m_1 - m_2$ is odd. The subspace spanned by $|m_{1}, m_{2},+-\rangle$ and $|m_{1}, m_{2},--\rangle$ will be referred to as $\mathscr{H}_{+-}$ and $\mathscr{H}_{--}$, respectively. The system is time-reversal invariant, and the chiral operator \( C \) maps states between \( \mathscr{H}_{+-} \) and \( \mathscr{H}_{--} \), so the Hamiltonians in these subspaces belong to the orthogonal class AI \cite{fan2017quantum,altland1997nonstandard,guhr1998random}.

Introducing the rescaled angular momentum vector $\widetilde{\boldsymbol{L}}^k=\boldsymbol{L}^k/(j+1/2)$, we ensure that in the limit $j\to\infty$, it satisfies $(\widetilde{\boldsymbol{L}}^k)^2=1$. The commutator then becomes $[\widetilde{L}_{x}^{k},\widetilde{L}_{y}^{k^{\prime}}]=i \hbar_{\text{eff}}\widetilde{L}_{z}^{k}\delta_{kk^{\prime}}$, with $\hbar_{\text{eff}} = {1}/{(j+1/2)}$ serving as the effective Planck constant. For simplicity, we omit the tilde in the rescaled angular momentum operators $\widetilde{L}_{x,y,z}^{k}$ and denote them by $L_{x,y,z}^{k}$ throughout the rest of the paper. The semiclassical limit is given by $\hbar_{\text{eff}} \rightarrow 0$, and the classical analogue of $H$ reads
\begin{equation}
    \mathcal{H}=(1+\lambda)(\mathcal{L}_z^1+\mathcal{L}_z^2)+4(1-\lambda)\mathcal{L}_x^1\mathcal{L}_x^2,
\end{equation}
 Here ${\cal L}_{x,y,z}^{k}$ are classical angular momentum variables, satisfying $({\cal L}_{x}^{k})^{2}+({\cal L}_{y}^{k})^{2}+({\cal L}_{z}^{k})^{2}=1$.
 The classical phase space is given by \(\mathbb{S}^2 \times \mathbb{S}^2\). Canonical coordinates \((z_k, \theta_k)\) can be introduced, using \(\mathcal{L}^k\) as $\mathcal{L}_x^k = \sqrt{1 - z_k^2} \cos \theta_k, \ 
\mathcal{L}_y^k = \sqrt{1 - z_k^2} \sin \theta_k, \
\mathcal{L}_z^k = z_k$. These expressions parametrize each two-dimensional sphere \(\mathbb{S}^2\) in terms of the canonical coordinates \((z_k,\theta_k)\), where \(z_k = \cos\phi_k\) represents the polar coordinate with \(\phi_k\in [0,\pi]\), and \(\theta_k \in [0,2\pi]\) is the azimuthal angle. 

In terms of canonical coordinates, the classical Hamiltonian is given by  
\begin{equation}  
    \mathcal{H} = (1+\lambda)(z_{1}+z_{2}) + 4(1-\lambda) \prod_{k=1}^{2} \sqrt{1 - z_{k}^{2}} \cos\theta_{k}.  
\end{equation}  
The energy of the classical Hamiltonian is bounded within the range \( [E_{\min}^\lambda, E_{\max}^\lambda] \), where  
$
E_{\min}^\lambda = \min{\mathcal{H}(\mathbf{z}_s, \mathbf{\theta}_s)}, \
E_{\max}^\lambda = \max{\mathcal{H}(\mathbf{z}_s, \mathbf{\theta}_s)}.
$
Here, \( (\mathbf{z}_s, \mathbf{\theta}_s) \) represent phase points where the gradient satisfies \( \nabla\mathcal{H} = 0 \).  
Due to the mirror symmetry of the system, it follows that $E_{\min}^\lambda + E_{\max}^\lambda = 0$. The maximum energy \( E_{\max}^\lambda \), as a function of \( \lambda \):
{\begin{equation}
    E_{\max}^\lambda =
    \begin{cases} 
        4(1-\lambda) + \frac{(1+\lambda)^2}{4(1-\lambda)}, & 0 \leq \lambda \leq \frac{3}{5}, \\ 
        2(1+\lambda), & \frac{3}{5} \leq \lambda \leq 1.
    \end{cases}
\end{equation}
}

\section{Chaos and ergodicity across the energy spectrum}\label{sec-spectrum}
{We begin by considering the average properties of the spectrum.
In particular, we study the averaged density of states and compare it with the semiclassical prediction.}

{In the quantum case, the averaged density of states can be defined as}
\begin{equation}\label{eq-dosq}
    \overline{\rho}(E)=\sum_{\alpha}\delta_{\epsilon}(E_{\alpha}-E) ,
\end{equation}
where $\delta_{\epsilon}(x)$ indicates the coarse-grained $\delta$ function
\begin{equation}\label{def-deltac}
    \delta_{\epsilon}(x)=\begin{cases}
\frac{1}{\epsilon} & |x|\le\frac{\epsilon}{2}\\
0 & |x|>\frac{\epsilon}{2}
\end{cases}.
\end{equation}
{The parameter $\epsilon$ is chosen to be sufficiently small compared to the spectral range of the system, yet still much larger than the mean level spacing. Note that other forms of the coarse-grained $\delta$ function, such as a Gaussian profile, can also be used in place of \eqref{def-deltac} without affecting the main results.
}

\begin{figure}
    \centering
    \includegraphics[width=1\linewidth]{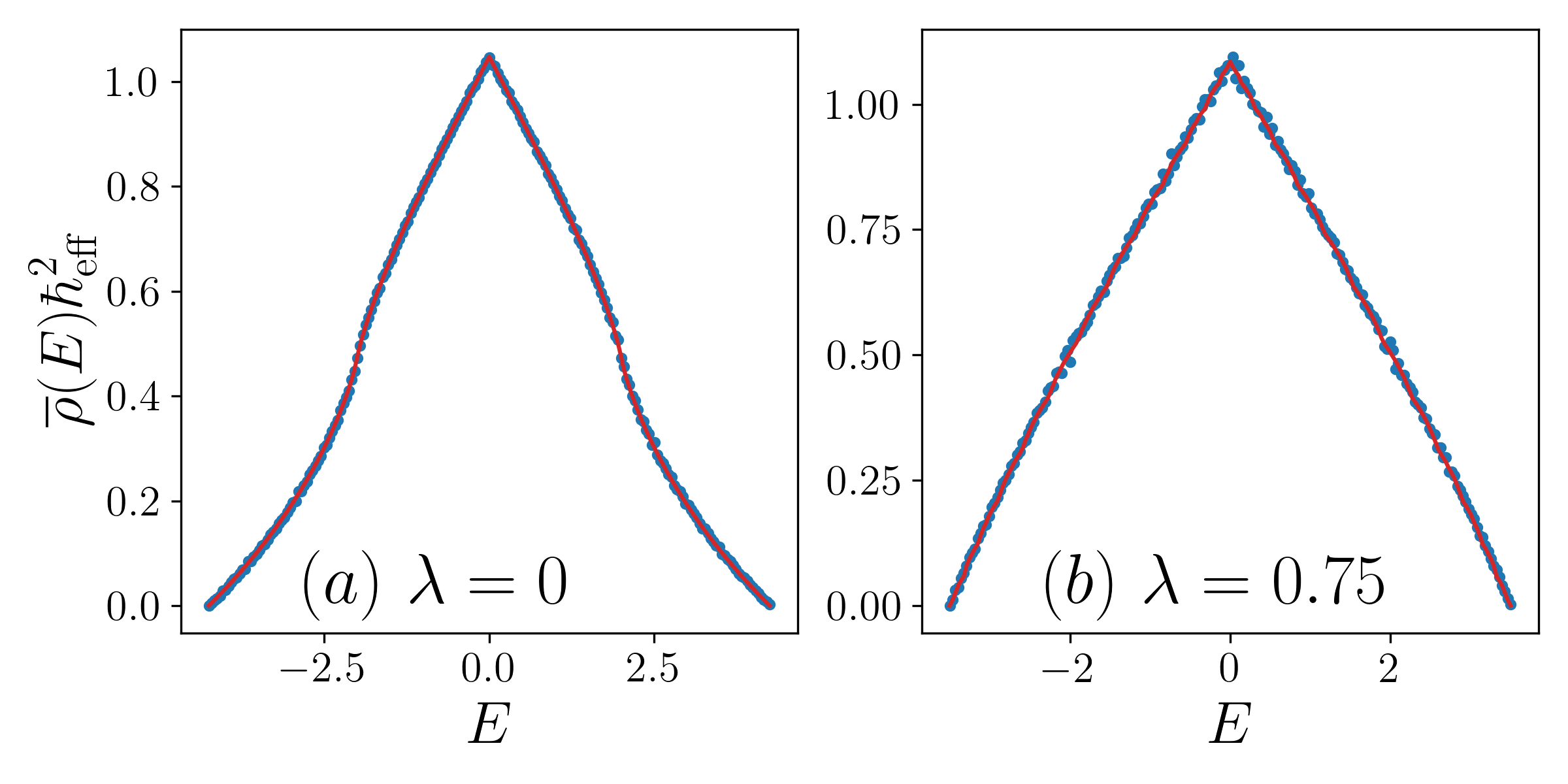}
    \caption{Averaged density of states: quantum $\hbar_{\text{eff}}^2\overline{\rho}(E)$ (blue circle) versus classical $\hbar_{\text{eff}}^2\rho_{\text{cl}}(E)$  (solid line) for (a) $\lambda=0$ and (b) $\lambda=0.75$. 
    In quantum case $j = 200$.
    $\overline{\rho} (E)$ is defined in Eq.~\eqref{eq-dosq}, and $\rho_{\text{cl}}(E)$ is defined in Eq.~\eqref{eq-rho-cl}.  Here we chose $\epsilon = 0.0425$ and the mean level spacing of the system $\overline{\delta e} \approx 10^{-4} $.}
    \label{Fig-dos}
\end{figure}

{The semiclassical density of states is given by the Thomas-Fermi expression \cite{brack2018semiclassical}:
\begin{equation}
    \rho_{\text{cl}}(E)=\frac{1}{(2\pi\hbar_\text{eff})^2}\int d\mathbf{z}d\mathbf{\theta}\ \delta(E - \mathcal{H}(\mathbf{z},\mathbf{\theta})) ,
\end{equation}
which is the classical phase space volume at energy $E$ divided by $(2\pi\hbar_\text{eff})^2$.
As shown in Appendix.~\ref{app-0}, in our model it reduces to}
\begin{equation}\label{eq-rho-cl}
    \rho_{\text{cl}}(E)=\frac{1}{(2\pi\hbar_\text{eff})^2}\int d\mathbf{z} \frac{8}{\xi}K(1-\frac{\eta^2}{\xi^2})\Theta(1-\frac{\eta^2}{\xi^2}).
\end{equation}
Here $K(a)$ is the complete elliptic integral of the first kind with parameter $a$, $\Theta(x)$ is the Heaviside step function, and
\begin{subequations}
\label{eq-integ-parameters}
   \begin{gather}
 \xi = 4(1-\lambda)\sqrt{1-z_{1}^{2}}\sqrt{1-z_{2}^{2}}\ , \\
\eta=E-(z_{1}+z_{2})(1+\lambda)\ .
\end{gather} 
\end{subequations}
{As a numerical check, in Fig.~\ref{Fig-dos} we compare the quantum averaged density of states $\overline{\rho}(E)$ with the semiclassical expression given in Eq.~\eqref{eq-rho-cl}, where good agreement is observed in both chaotic and (near-)integrable cases.
To quantify the agreement between the two distributions, we employ
\begin{equation}
\delta_{\rho}=\frac{\int_{E_{\text{min}}}^{E_{\text{max}}}|\overline{\rho}(E)-\rho_{\text{cl}}(E)|dE}{\int_{E_{\text{min}}}^{E_{\text{max}}}\rho_{\text{cl}}(E)dE}\ ,
\end{equation}
where, in numerical calculations, the integrals are approximated by sums over bins of width $\epsilon$.
For this measure, we obtain $\delta_{\rho} = 0.0045$ for $\lambda = 0.0$ and $\delta_{\rho} = 0.011$ for $\lambda = 0.75$ . We note that $\overline{\rho}(E)$ denotes the averaged density of states of the full Hilbert space (dimension $(2j+1)^2$). In practice, we restrict our analysis to the subspace ${\mathscr H}_{+-}$ and approximate it by $\overline{\rho}(E) = 4\overline{\rho}_{+-}(E)$, where  $\overline{\rho}_{+-}(E)$ indicates the averaged density of states in ${\mathscr H}_{+-}$.
}

\begin{figure}
    \centering
    \includegraphics[width=1\linewidth]{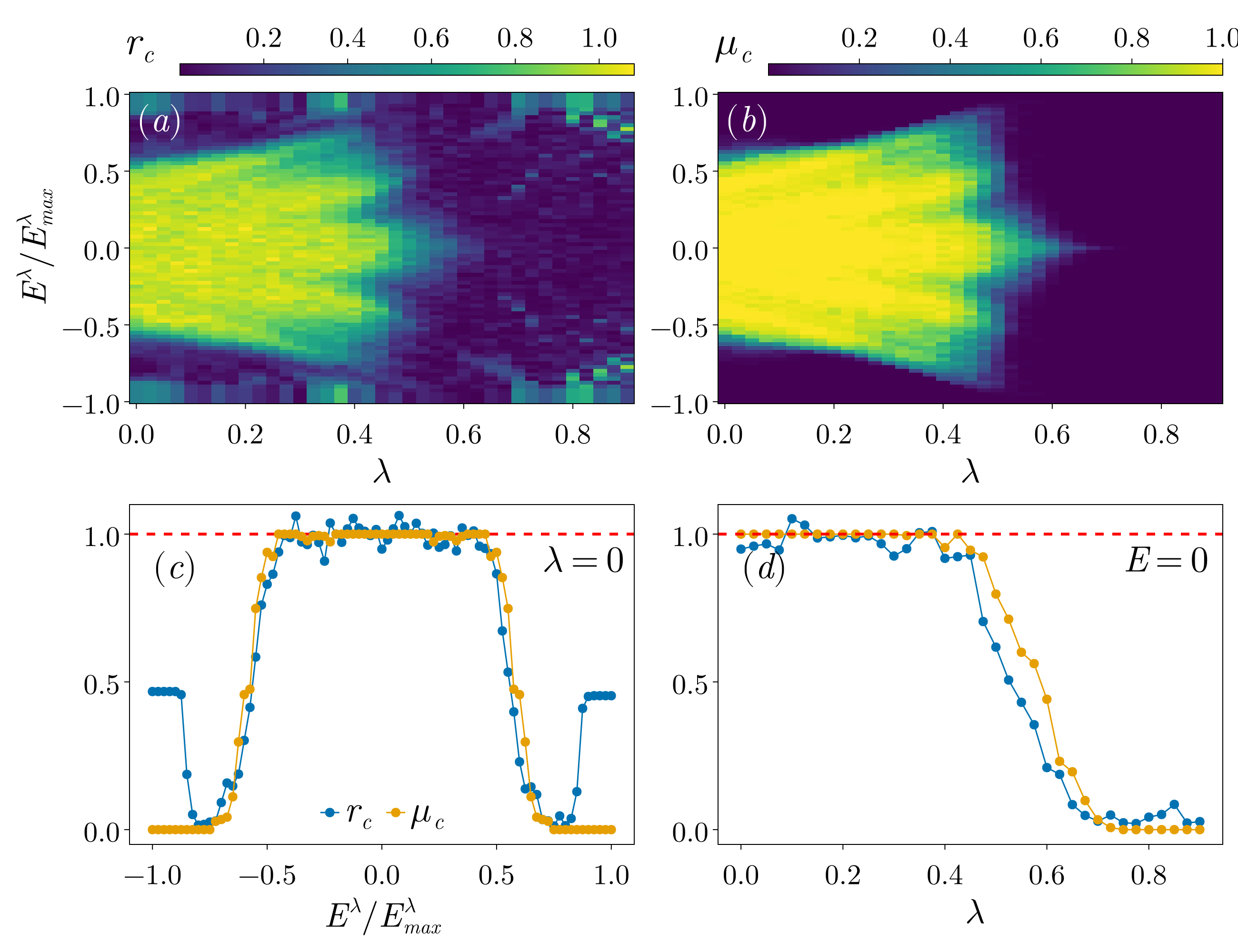}
    \caption{Chaos and ergodicity across the energy spectrum, probed by (a)  $r_c$ (given in Eq.~\eqref{eq-dsp}) the normalized mean spacing ratio, and (b) classical indicator $\mu_c$, the fraction of chaotic region on energy surface.  In (a) we use $500$ eigenvalues and average over $10$ different $j$ around $j = 200$. In (b) we sample 200 random points on each classical energy shell to approximate \(\mu_c\) from Eq.~\eqref{eq-deltac}, where the characteristic function \(\chi_c = 1\) for SALI \(\leq 10^{-8}\) at \(t = 180\), and \(0\) otherwise. Panels (c) and (d) display transitions of $r_c$ and $\mu_c$ as functions of $E$ and $\lambda$, with $\lambda=0$ and $E=0$, respectively.}
    \label{Fig-DF}
\end{figure}

Next, we examine the fluctuation properties of the spectrum by analyzing the ratio of consecutive energy levels \cite{PhysRevB.75.155111-r-huse}. It is defined as
\begin{equation}
r_{\alpha}=\frac{\min(s_{\alpha},s_{\alpha-1})}{\max(s_{\alpha},s_{\alpha-1})},
\end{equation}
where $s_{\alpha} = E_{\alpha+1}-E_\alpha$ is the nearest-neighbor spacing between energy levels. In particular, we consider the mean value of $r_\alpha$ in the energy window of interest, denoted by $\langle r \rangle$. {In integrable systems, where eigenvalues are typically uncorrelated, the level spacings obey Poisson statistics, and the mean spacing ratio  \( \langle r \rangle_P = 2\ln 2 -1 \). In fully chaotic systems, on the other hand, eigenvalues exhibit correlations and the level spacings follow GOE statistics, for which \( \langle r\rangle_{\text{GOE}}  \approx 0.53 \)}. We define the normalized mean spacing ratio, \( r_c \), as an indicator of quantum chaos:  
\begin{equation}\label{eq-dsp}
r_c=\frac{\left|\langle r\rangle-\langle r\rangle_{\text{P}}\right| }{\langle r\rangle_{\text{GOE}}-\langle r\rangle_{\text{P}}}.
\end{equation}  
Clearly, \( r_c = 0 \) for an integrable system and \( r_c = 1 \) for a fully chaotic system.

For comparison, we also introduce a classical {indicator} of chaos.
Let $\chi_c(\mathbf{z},\mathbf{\theta})$ denote the characteristic function of the
chaotic component in the classical phase space, which takes the value of 1 in the chaotic region and zero otherwise. The chaotic fraction $\mu_{\text{c}}$
is defined as the relative Liouville volume of the chaotic region within the classical phase space on the energy shell as
\begin{align}\label{eq-deltac}
    \mu_c = \frac{\int d\mathbf{z}d\mathbf{\theta} \, \chi_c(\mathbf{z},\mathbf{\theta}) \delta(E - \mathcal{H}(\mathbf{z},\mathbf{\theta}))}{\int d\mathbf{z}d\mathbf{\theta} \, \delta(E - \mathcal{H}(\mathbf{z},\mathbf{\theta}))}.
\end{align}
We employ the smaller alignment index (SALI) \cite{skokos2016chaos}
 to evaluate the characteristic function $\chi_c(\mathbf{z},\mathbf{\theta})$. This approach evaluates deviation vectors from a given orbit, based on the equations of the tangent map
{derived by linearizing the equations of motion}  $\dot{\mathbf{w}} = \Omega \cdot \nabla^2 {\cal H} (\mathbf{x}) \cdot \mathbf{w}$  with $\Omega = \begin{bmatrix}  0 & -I_2\\
    I_2 & 0
    \end{bmatrix}$,
where $\mathbf{x}=(\mathbf{z},\mathbf{\theta})$ and the deviation vector $\mathbf{w}=\delta \mathbf{x}$, $\nabla^2 {\cal H}(\mathbf{x})$ is the Hessian matrix, $I_2$ being the two-dimensional identity matrix. For numerical evaluation, we employ Monte Carlo sampling with 200 random points from the classical energy shell for each $(E, \lambda)$ to approximate $\mu_c$, where the characteristic function $\chi_c = 1$ for SALI $\leq 10^{-8}$ at $t = 180$, and $0$ otherwise. We leave more details about SALI and the criteria of $\chi_c$ in Appendix \ref{app-1}.

The mean spacing ratio, shown in Fig.~\ref{Fig-DF}(a), is compared with the classical chaos indicator \(\mu_c\), shown in Fig.~\ref{Fig-DF}(b). The latter quantifies the fraction of the chaotic region as defined in Eq.~\eqref{eq-deltac}. Both exhibit a similar general pattern across the energy spectrum and for various values of $\lambda$, as also demonstrated by random matrix theory in an extended Rosenzweig-Porter approach \cite{yan2025spacing}. 
{Small anomalous regions appear near \(E/E_{\text{max}}=\pm 1\) in Fig.~\ref{Fig-DF}(c), where the classical dynamics are regular and, according to the Berry--Tabor conjecture, one would generally expect \(r_c=0\). These deviations are likely caused by the low density of states and the limited Hilbert-space dimension, which leave too few eigenvalues within the energy window for robust statistics. They should therefore diminish with increasing system size.} In the following, we study the statistical properties of observables and the emergence of random matrix theory universality, in the chaotic region confirmed by both the spacing ratio and chaotic fraction in the classical limit, with the case in the (near-)integrable region as a comparison.

\section{Statistical properties of observables}\label{sec-observable}

\subsection{Conventional ETH}\label{sec-eth-old}
In few-body systems with a classical limit, based on previous studies \cite{offeth_PhysRevE.60.1630,offeth_PhysRevLett.75.2300,offeth_Wilkinson_1987,feingold1986distribution,Srednicki_1996}, we propose the following form of ETH ansatz,
\begin{subequations}
\label{eq-ETH-new}
\begin{gather}
    {\cal O}_{\alpha\beta}={\cal O}(\overline{E})\delta_{\alpha\beta}+\rho^{-1/2}(\overline{E})f(\overline{E},\omega)r_{\alpha\beta} , \label{eq-ETH-new0} \\
f(\overline{E},\omega)=\hbar_{\text{eff}}^{-1/2}g(\overline{E},\omega/\hbar_{\text{eff}}) . \label{eq-ETH-new1}
\end{gather}
\end{subequations}
Both ${\cal O}(\overline{E})$ and $g(\overline{E},\omega)$ are of order $1$ and smooth functions of their argument. {The density of states $\rho(\overline{E})$ scales as $\rho(\overline{E})\sim \hbar_{\text{eff}}^{-f}$, where $f$ is 
the number of degrees of freedom, which equals 2 in our model.} Similarly to the many-body case, $r_{\alpha\beta} = r^*_{\beta\alpha}$ is a factor whose real and imaginary parts are assumed to be pseudo-random variables with zero mean and unit variance.
{An implication of Eq.~\eqref{eq-ETH-new} is that} the fluctuation of the diagonal part scales as 
\begin{equation}\label{eq-ddeth}
    \delta {\cal O}_{\alpha\alpha} \sim \hbar_{\text{eff}}^{-1/2}\rho^{-1/2}(\overline{E})
\sim\hbar_{\text{eff}}^{1/2} .
\end{equation}
In the Appendix.~\ref{sec-appendixb}, {we provide a detailed discussion} of the scaling of $f(\overline{E},\omega)$ with $\hbar_{\text{eff}}$ given in Eq.~\eqref{eq-ETH-new1}.

{Through numerical simulations, some features consistent with the ETH ansatz in few-body systems have been observed \cite{David22}, including, e.g., the smoothness of ${\cal O}(\overline{E})$ and the Gaussianity of $r_{\alpha\beta}$. }
However, many aspects of ETH remain to be verified. {A particularly important aspect is the $\hbar_{\text{eff}}$ dependence of both the envelope function $g(\overline{E},\omega)$ and the fluctuations of the diagonal elements.} In this section, we study the question in the Feingold-Peres model, where we consider the following two observables
\begin{equation}
    {\cal A} = L^1_z + L^2_z,\ {\cal B} = L^1_z - L^2_z .
\end{equation}
It is easy to see that $\mathcal{A}$ and $\mathcal{B}$ are symmetric or antisymmetric, respectively, under the exchange of two angular momenta. Accordingly, in the remainder of this study, $\mathcal{A}$ is analyzed within the subspace $\mathscr{H}_{+-}$, and $\mathcal{B}$ within $\mathscr{H}_{+-} \oplus \mathscr{H}_{--}$.

\begin{figure}[]
	\includegraphics[width=0.85\linewidth]{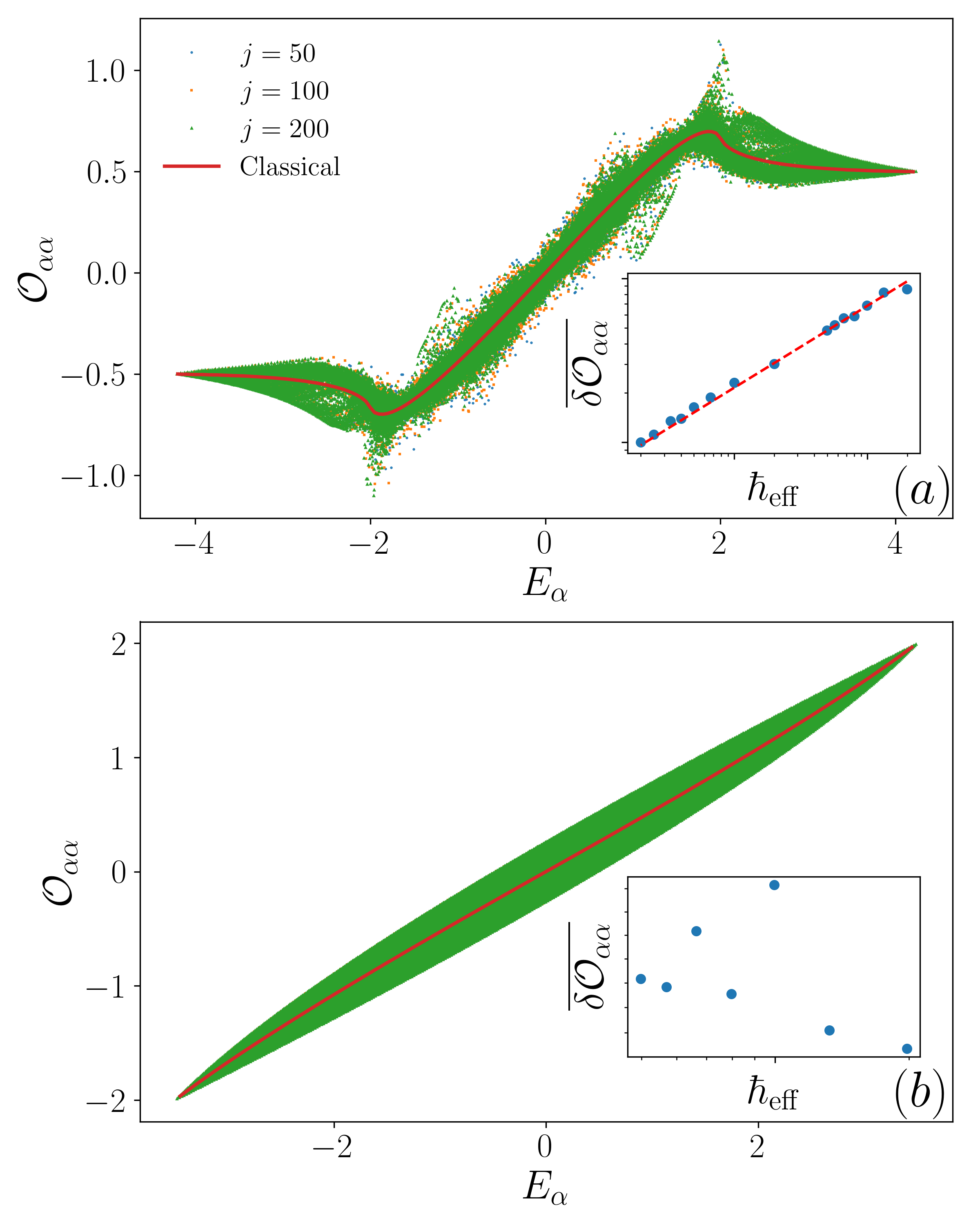}
 \caption{Diagonal elements: ${\cal O}_{\alpha \alpha}$ versus $E_\alpha$ for operator ${\cal A} = L^1_z + L^2_z $ for (a) $\lambda = 0$ and (b) $\lambda = 0.75$ .
 Inset: the average nearest-neighbor difference of diagonal elements $\overline{\delta {\cal O}_{\alpha\alpha}}$, where the average is taken in the energy window $E_\alpha \in [-0.05, 0.05]$ for $j\ge 200$. In the inset of (a) $\overline{\delta {\cal O}_{\alpha\alpha}}$ is also shown for $j = 500,1000,\cdots, 5000$, where the average is taken over $2000$ different $\alpha$ ($1000$ for $j \ge 4000$) in the middle of  the spectrum.
 The dashed line indicates the scaling $\propto \hbar_{\text{eff}}^{1/2}.$ 
 }\label{Fig-DiagETH}
\end{figure}

The envelope function of the diagonal elements is defined as
\begin{equation}
{\cal O}(E)=\frac{1}{\sum_{\alpha}\delta_{\epsilon}(E_{\alpha}-E)}\sum{\cal O}_{\alpha\alpha}\delta_{\epsilon}(E_{\alpha}-E) .
\end{equation}
{Similar to Eq.~\eqref{def-deltac}, $\epsilon$ should also be chosen such that it is much smaller than the spectral range, but still much larger than the mean level spacing of the system. } 
In the semiclassical limit $\hbar_{\text{eff}}\rightarrow 0$, a classical expression of ${\cal O}(E)$ can be derived \cite{wang2022semiclassical,Srednicki_1999}, which in case of our model, reads
  \begin{equation} \label{eq-diag-cl}
      {\cal O}^{\text{cl}}(E)=\frac{1}{S(E)}\int d\boldsymbol{z}d\boldsymbol{\theta}{\cal O}^{\text{cl}}(\boldsymbol{z},\boldsymbol{\theta})\delta({\cal H}(\boldsymbol{z},\boldsymbol{\theta})-E) ,
  \end{equation}
where
\begin{equation}
      S(E)=\int d\boldsymbol{z}d\boldsymbol{\theta}\delta({\cal H}(\boldsymbol{z},\boldsymbol{\theta})-E)\ .
\end{equation} 
The classical analogue of ${\cal A}$ and ${\cal B}$ reads
\begin{equation}
{\cal A}^{\text{cl}}(\boldsymbol{z},\boldsymbol{\theta}) = {\cal L}^1_z + {\cal L}^2_z,\quad {\cal B}^{\text{cl}}(\boldsymbol{z},\boldsymbol{\theta}) = {\cal L}^1_z - {\cal L}^2_z .
\end{equation}
Similar to Eq.~\eqref{eq-rho-cl}, we have
\begin{align}
 {\cal A}^{\text{cl}}(E)=&\frac{1}{S(E)}\int dz_{1}dz_{2}(z_{1}+z_{2})\frac{8}{\xi}K(\mu)\Theta(\mu) \nonumber \\
{\cal B}^{\text{cl}}(E)=&\frac{1}{S(E)}\int dz_{1}dz_{2}(z_{1}-z_{2})\frac{8}{\xi}K(\mu)\Theta(\mu),
\end{align}
where $S(E)=\int dz_{1}dz_{2}\frac{8}{\xi}K(\mu)\Theta(\mu)$,
$\mu=1-\eta^2/\xi^2$, and $\eta$ and $\xi$ are defined in Eq.~\eqref{eq-integ-parameters}. It is straightforward to see that ${\cal B}^{\text{cl}}(E) = 0$, as this follows directly from the antisymmetry of ${\cal B}$ under the exchange of two angular momenta.


Diagonal elements of ${\cal A}$ are shown in Fig.~\ref{Fig-DiagETH} for different $j$. Results are also compared with the classical expression given in Eq.~\eqref{eq-diag-cl} . In both chaotic ($\lambda = 0$) and (near-)integrable cases ($\lambda = 0.75$) , ${\cal O}_{\alpha\alpha}$
{exhibit noticeable fluctuations, while their averaged profile is close to the classical prediction}, and a roughly linear ${\cal O}(E)\propto E$ is observed at $E\approx 0$. To characterize the fluctuation of diagonal parts, 
we consider 
{\begin{equation}\label{eq-dO}
\delta{\cal O}_{\alpha\alpha}=|{\cal O}_{\alpha\alpha}-{\cal O}_{\alpha-1,\alpha-1}|.
\end{equation}}
More specifically, we focus on its average in a small energy window around $E=0$ denoted by $\overline{\delta{\cal O}_{\alpha\alpha}}$.  
For the chaotic case $\lambda = 0$, we find $\overline{\delta{\cal O}_{\alpha\alpha}} \propto \hbar_{\text{eff}}^{1/2}$. It agrees with the prediction in Eq.~\eqref{eq-ddeth}, consistent with the ETH ansatz for few-body systems given in Eq.~\eqref{eq-ETH-new}. This indicates $\overline{\delta{\cal O}_{\alpha\alpha}}$ will vanish in the semiclassical limit $\hbar_{\text{eff}} \rightarrow 0$. In contrast, for the (near-)integrable case $\lambda = 0.75$, $\overline{\delta{\cal O}_{\alpha\alpha}}$ does not decay with decreasing $\hbar_{\text{eff}}$, at least for the value of $\hbar_{\text{eff}}$ considered here. This suggests {a violation} of ETH in the (near-)integrable case.

\begin{figure}[]
	\includegraphics[width=1.0\linewidth]{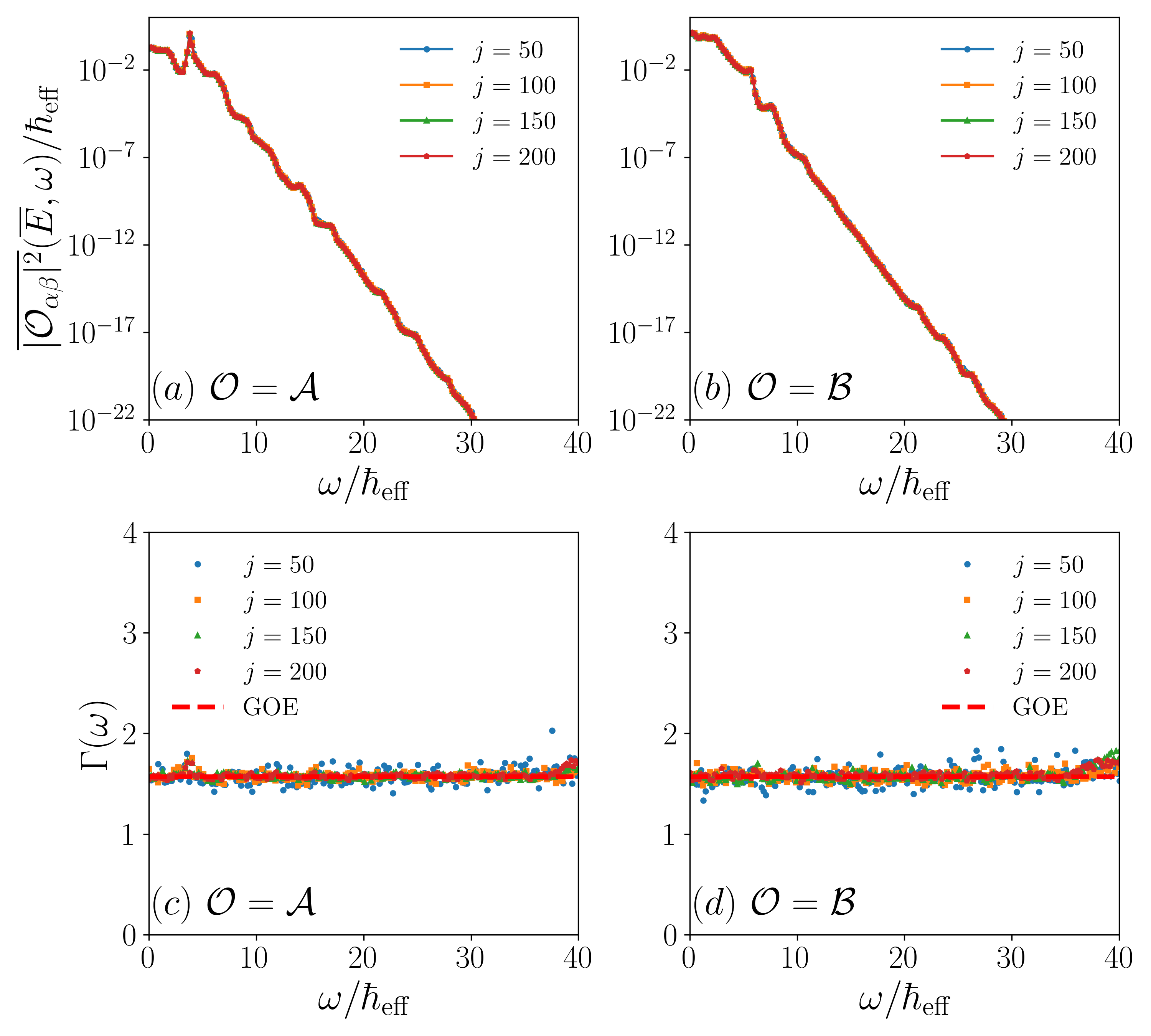}
 \caption{Statistical properties of off-diagonal elements 
 ${\cal O}_{\alpha \beta}$ in chaotic case $\lambda = 0.0$.
$\overline{|{\cal O}_{\alpha\beta}|^{2}}(\overline{E},\omega)/\hbar_{\text{eff}}$ versus $\omega/\hbar_{\text{eff}}$ for operator (a) ${\cal A}$ and (b) ${\cal B}$ for $\overline{E} \approx 0$ .
 Indicator of Gaussianity of  ${\cal O}_{\alpha \beta}$, $\Gamma(\omega)$ versus $\omega/\hbar_{\text{eff}}$ for operator (c) ${\cal A}$ and (d) ${\cal B}$ .
 The dashed line indicates the prediction of GOE, $\Gamma = \frac{\pi}{2}$ .}\label{Fig-FW}
\end{figure}

\begin{figure}[]
	\includegraphics[width=1.0\linewidth]{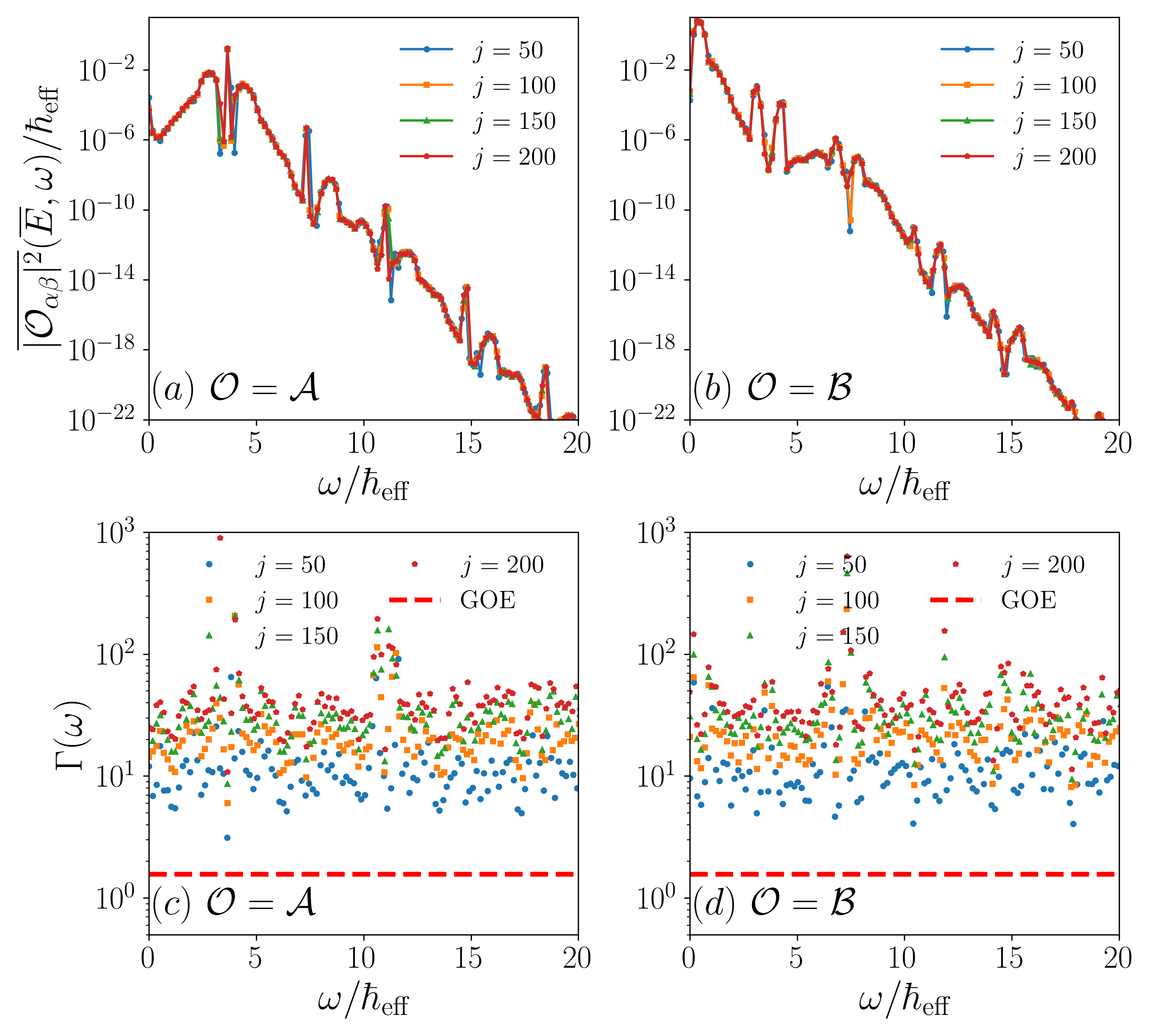}
 \caption{Statistical properties of off-diagonal elements 
 ${\cal O}_{\alpha \beta}$. Similar to Fig.~\ref{Fig-FW} but for (near-)integrable case $\lambda = 0.75$.}\label{Fig-FW-Int}
\end{figure}

The envelope function of off-diagonal elements $f(\overline{E},\omega)$ can also be studied by semiclassical analysis \cite{offeth_PhysRevE.61.R2180,offeth_Wilkinson_1987}. 
However, in generic cases, the semiclassical expression is very involved, which we do not discuss here. We primarily rely on numerical simulations. To this end, we consider
\begin{gather}
\overline{|{\cal O}_{\alpha\beta}|^{2}}(\overline{E},\omega)=\frac{1}{N_{\omega}}\sum_{\alpha\neq\beta}|{\cal O}_{\alpha\beta}|^{2}\delta_{\epsilon}(E_{\alpha}-E_{\beta}-\omega) \nonumber \\
\times\ \delta_{\epsilon^{\prime}}(\frac{E_{\alpha}+E_{\beta}}{2}-\overline{E}) ,
\end{gather}
and 
\begin{gather}
\overline{|{\cal O}_{\alpha\beta}|}(\overline{E},\omega)=\frac{1}{N_{\omega}}\sum_{\alpha\neq\beta}|{\cal O}_{\alpha\beta}|\delta_{\epsilon}(E_{\alpha}-E_{\beta}-\omega) \\ \nonumber
\times \ \delta_{\epsilon^{\prime}}(\frac{E_{\alpha}+E_{\beta}}{2}-\overline{E}),
\end{gather}
where 
\begin{equation}
N_{\omega}=\sum_{\alpha\neq\beta}\delta_{\epsilon}(E_{\alpha}-E_{\beta}-\omega)\delta_{\epsilon^{\prime}}(\frac{E_{\alpha}+E_{\beta}}{2}-\overline{E}) .
\end{equation}
Here, both $\epsilon$ and $\epsilon'$ should be sufficiently small compared to the whole energy range, but still much larger than the mean level spacing of the system.
According to Eq.~\eqref{eq-ETH-new},
the envelope function $f(\overline{E},\omega)$ is related to $\overline{|{\cal O}_{\alpha\beta}|^{2}}(\overline{E},\omega)$ as
\begin{equation}\label{eq-f-O}
    f^{2}(\overline{E},\omega)=\rho(\overline{E})\overline{|{\cal O}_{\alpha\beta}|^{2}}(\overline{E},\omega) .
\end{equation}
Given that $\rho(\overline{E})\sim \hbar_{\text{eff}}^{-2}$, {one has}
\begin{equation}\label{eq-f-O2}
    f^{2}(\overline{E},\omega)\sim\hbar_{\text{eff}}^{-2}\overline{|{\cal O}_{\alpha\beta}|^{2}}(\overline{E},\omega) .
\end{equation}
{The Gaussianity can be characterized by the ratio
\begin{equation}\label{eq-Gamma}
    \Gamma(\overline{E},\omega)=\frac{\overline{|{\cal O}_{\alpha\beta}|^{2}}(\overline{E},\omega)}{[\overline{|{\cal O}_{\alpha\beta}|}(\overline{E},\omega)]^{2}},
\end{equation}
for which \(\Gamma(\overline{E},\omega)=\pi/2\) when \({\cal O}_{\alpha\beta}\) follows a Gaussian distribution.}
In our simulations, we fix $\overline{E} = 0$, and denote $\Gamma(\overline{E},\omega)$ by $\Gamma(\omega)$ for simplicity.

{Figure~\ref{Fig-FW}(a) and (b) show the results in the chaotic case $\lambda = 0$ for the two observables 
${\cal A}$ and ${\cal B}$. A smooth structure is observed, in agreement with ETH.} Additionally, an approximate data collapse is seen when plotting 
$\overline{|{\cal O}_{\alpha\beta}|^{2}}(\overline{E},\omega)/\hbar_{\text{eff}}$ as a function of $\omega/\hbar_{\text{eff}}$. This suggests the scaling 
\begin{equation}\label{eq-oab-scaling}
\overline{|{\cal O}_{\alpha\beta}|^{2}}(\overline{E},\omega)\sim\hbar_{\text{eff}} g^2(\overline{E},\omega/ \hbar_{\text{eff}}) ,
\end{equation}
where $g$ is a smooth function that does not scale with $\hbar_{\text{eff}}$.
Combining Eq.~\eqref{eq-f-O2} with Eq.~\eqref{eq-oab-scaling}, one obtains
\begin{equation}
f(\overline{E},\omega)\sim \hbar^{-1/2}_{\text{eff}}g(\overline{E},\omega /\hbar_{\text{eff}}),
\end{equation}
which agrees with the ETH ansatz given in Eq.~\eqref{eq-ETH-new}.
{Furthermore, an exponential decay is observed for both observables, similar to the results in many-body systems \cite{PhysRevE.100.062134-ETH-Lev,PhysRevE.102.042127}.}

To check the Gaussianity of ${\cal O}_{\alpha\beta}$, we show the ratio $\Gamma(\omega)$ for different $\omega$. In the chaotic case $\lambda = 0.0$, for almost all frequencies $\omega$ considered here, 
$\Gamma(\omega)$ closely follows the RMT prediction $\Gamma(\omega)=\frac{\pi}{2}$. This suggests that ${\cal O}_{\alpha \beta}$ indeed follows a Gaussian distribution in the chaotic regime. In contrast, for (near-)integrable case $\lambda = 0.75$, as shown in Fig.~\ref{Fig-FW-Int}, fluctuations of $\overline{|{\cal O}_{\alpha\beta}|^{2}}(\overline{E},\omega)$ are visible, but the general scaling remains consistent with Eq.~\eqref{eq-oab-scaling}. However, unlike in the chaotic case, the distribution of ${\cal O}_{\alpha\beta}$ is far from Gaussian, as $\Gamma(\omega)$ deviates significantly from the RMT prediction $\Gamma(\omega)=\frac{\pi}{2}$ for all considered $\omega$ (Fig.~\ref{Fig-FW-Int} (c)(d)). 

In summary, the numerical results in this section support the ETH ansatz proposed in Eq.~\eqref{eq-ETH-new} for few-body systems with a chaotic classical limit, while deviations are found in (near-)integrable systems.

\begin{figure}[t]
	\includegraphics[width=1.0\linewidth]{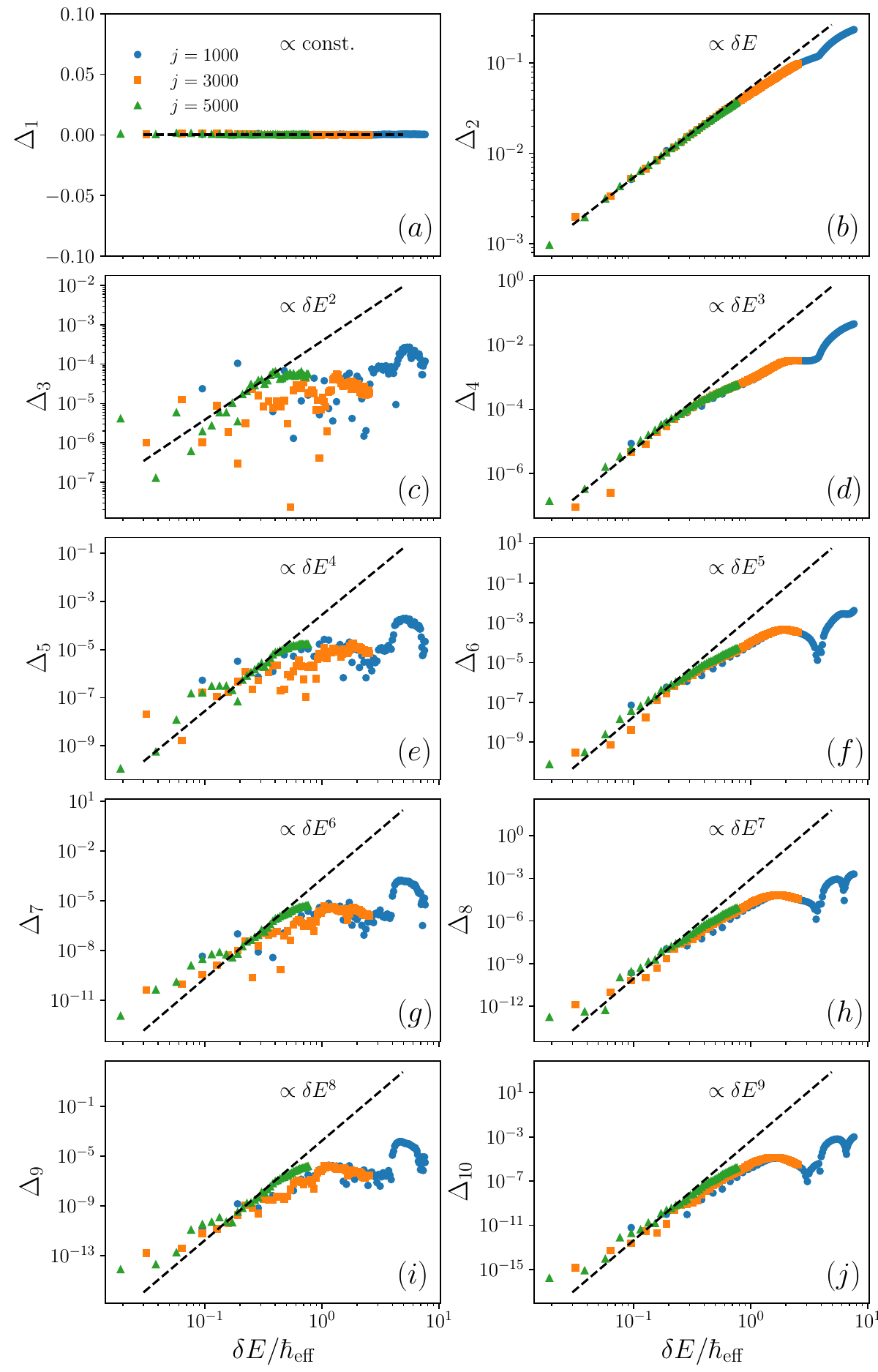}
 \caption{Indicator {for UIE description} of microcanonically truncated operators (chaotic case $\lambda = 0.0$): free cumulants $\Delta_k$ versus $\delta E / \hbar_{\text{eff}}$ for operator ${\cal A}$. {As a guide to the eye, the inclined dashed lines (black)  indicate the scaling $\Delta_k \propto \delta E^{k-1}$.} Here and in the following figures, the absolute values of the free cumulants are plotted.
 }\label{Fig-FC}
\end{figure}

\begin{figure}[t]
	\includegraphics[width=1.0\linewidth]{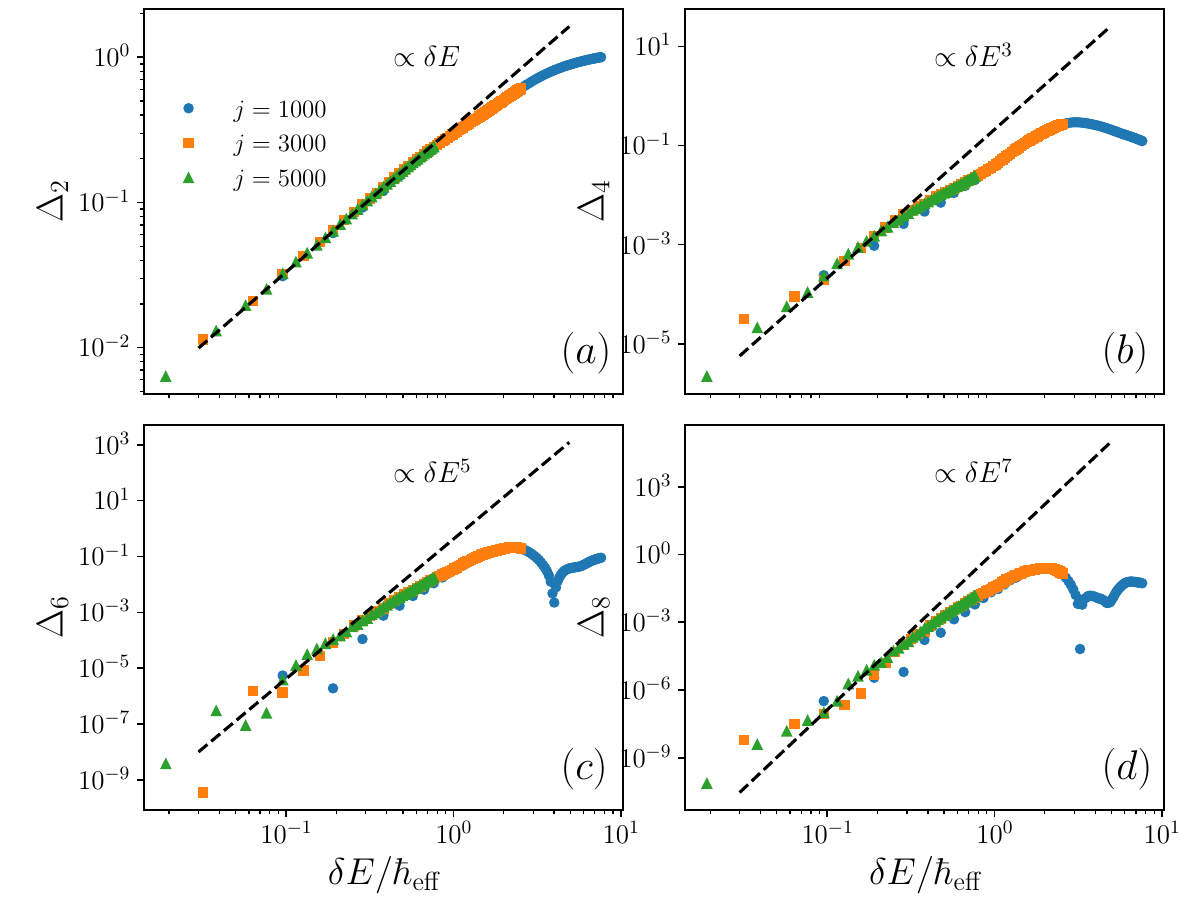}
 \caption{Indicator {for UIE description} of microcanonically truncated operators (chaotic case $\lambda = 0.0$): similar to Fig.~\ref{Fig-FC}, but for operator ${\cal B}$. Only even free cumulants are shown here since all odd free cumulants are $0$.
 }\label{Fig-FCB}
\end{figure}

\begin{figure}[t]
	\includegraphics[width=1.0\linewidth]{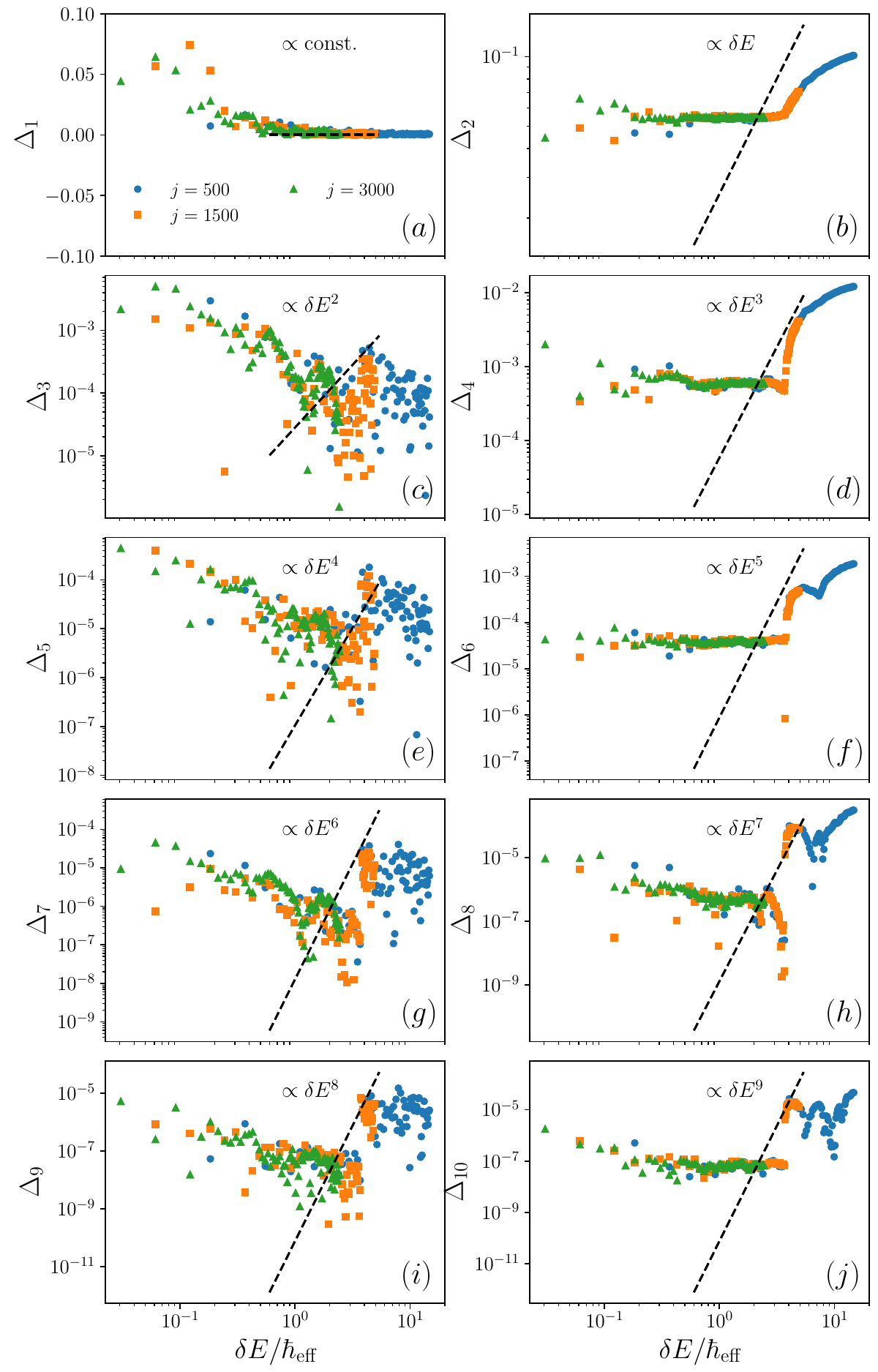}
  \caption{Indicator {for UIE description} of microcanonically truncated operators ((near-)integrable case $\lambda = 0.75$) : Free cumulants $\Delta_k$ versus $\delta E / \hbar_{\text{eff}}$ for operator ${\cal A}$. As a guide to the eye, the inclined dashed lines (black) indicate  $\Delta_k \propto \delta E^{k-1}$.
 }\label{Fig-FC-Int}
\end{figure}

\begin{figure}[t]
	\includegraphics[width=1.0\linewidth]{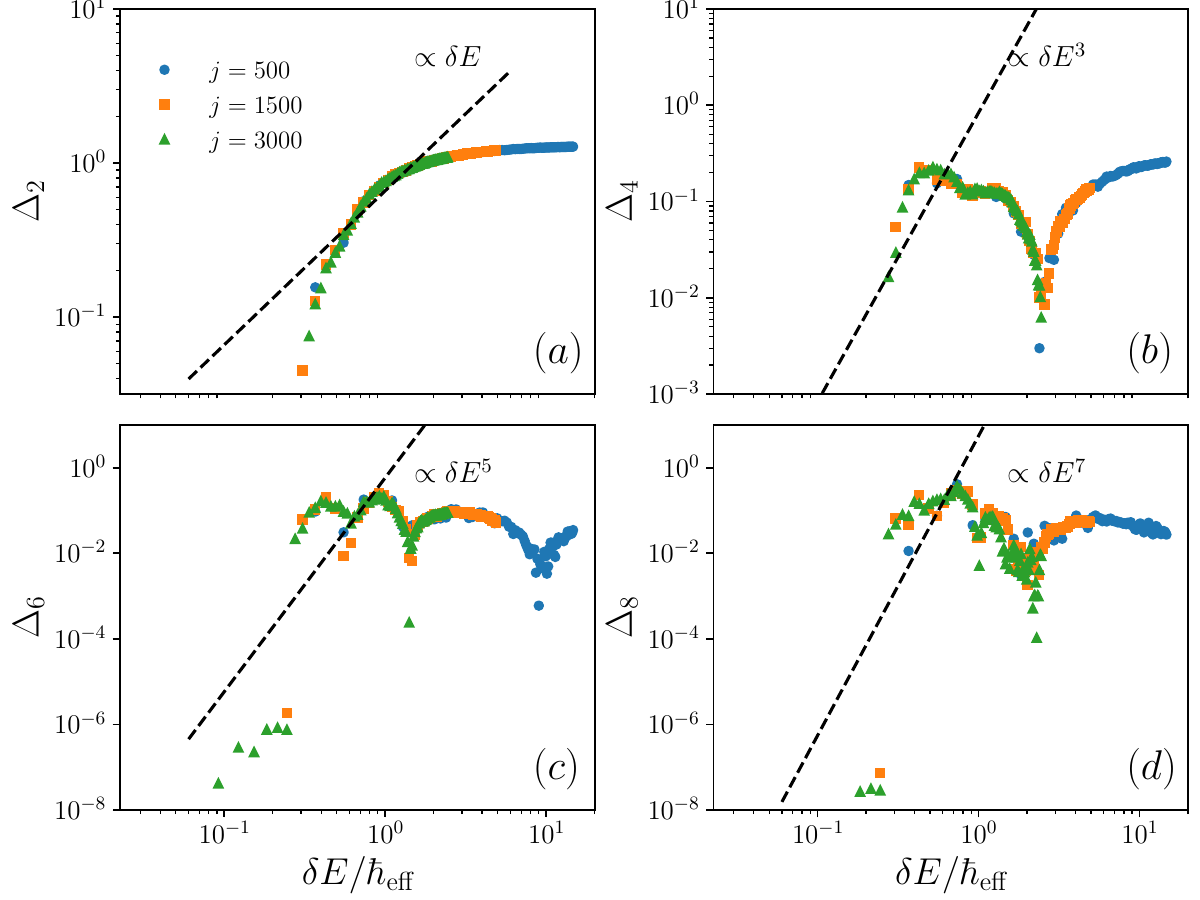}
 \caption{Indicator {for UIE description} of microcanonically truncated operators
((near-)integrable case $\lambda = 0.75$): similar to Fig.~\ref{Fig-FC-Int}, but for operator ${\cal B}$ . Only even free cumulants are shown here since all odd free cumulants are $0$.
 }\label{Fig-FCB-Int}
\end{figure}

\subsection{{Beyond conventional ETH: UIE description of microcanonically truncated operators}}\label{sec-eth-new}

{Let us consider a truncated operator
${\cal O}_{\Delta E}=P_{\Delta E}{\cal O}P_{\Delta E}$, where
\begin{equation}\label{eq-PDE}
P_{\Delta E}=\sum_{|E_{\alpha}-E_{0}|\le\Delta E/2}|\alpha\rangle\langle\alpha|
\end{equation}
is the projection operator onto a microcanonical energy window.
Motivated by the local unitary invariance of observable matrix-element statistics, the following picture of RMT universality has been conjectured\cite{wang2023emergence}: below a sufficiently small energy scale $\Delta E_U$, the truncated operator ${\cal O}_{\Delta E_U}$ can be regarded as a sample from a unitarily invariant ensemble (UIE),
\begin{equation}\label{eq-uie}
{\cal O}_{\Delta E_U} \sim {\cal O}_{U}=U{\cal O}^{\star}U^{\dagger},
\end{equation}
where $U$ is a Haar-random unitary (or orthogonal) operator and ${\cal O}^{\star}$ is a fixed, operator-specific matrix, which can be chosen to be diagonal without loss of generality.
}

To characterize the UIE, we consider submatrices of ${\cal O}_{\Delta E_U}$ obtained by projecting it onto an even smaller energy window, yielding
${\cal O}_{\delta E} = P_{\delta E} {\cal O}_{\Delta E_U}P_{\delta E}$.
{If the UIE description in Eq.~\eqref{eq-uie} applies, spectral properties of ${\cal O}_{\delta E}$ are uniquely fixed by the spectrum of ${\cal O}_{\Delta E_U}$ and the ratio $d_{\delta E}/d_{\Delta E_U}$.}
 For example, for sufficiently large $d_{\delta E}$ the $k$-th free cumulant of ${\cal O}_{\delta E}$ is given by \cite{wang2023emergence}
\begin{equation}\label{eq-fca}
    \Delta_{k}(\delta E)=\left(\frac{d_{\delta E}}{d_{U}}\right)^{k-1}\Delta_k^U,\ \text{for\ } \delta E \le \Delta E_U ,
\end{equation}
where $\Delta^U_k$ denotes the $k$-th free cumulant of ${\cal O}_{\Delta E_U}$ and $d_U \equiv d_{\Delta E_U}$. 
{Here, the free cumulants of an operator ${\cal O}$ are defined in terms of its moments ${\cal M}_k \equiv \frac{1}{d}\mathrm{Tr}[{\cal O}^k]$ \cite{PhysRevLett.129.170603}, where $d$ is the dimension of the corresponding Hilbert subspace, and are determined recursively via}
\begin{equation}\label{eq-fc}
\Delta_{k} = {\cal M}_{k}-\sum_{j=1}^{k-1}\Delta_{j}\sum_{a_{1}+a_{2}+\cdots a_{j}=k-j}{\cal M}_{a_{1}}\cdots{\cal M}_{a_{j}}\ .
\end{equation}
{In Appendix.~\ref{apx-fc}, we present details on the derivation of Eq.~\eqref{eq-fca}.}

If $\Delta E_U$ is sufficiently small, the density of states within the microcanonical window can be regarded as constant. It then follows that $d_{\delta E}/d_{U}=\delta E/\Delta E_{U}$ which reduces Eq.~\eqref{eq-fca} to the following relations
\begin{equation}\label{eq-Deltak-DE}
\Delta_{k}(\delta E)=\left(\frac{\delta E}{\Delta E_{U}}\right)^{k-1}\Delta_{k}^{U} \propto \delta E^{k-1} .
\end{equation}
{The scaling relations in Eq.~\eqref{eq-Deltak-DE} thus provide an infinite set of necessary conditions for the truncated operator ${\cal O}_{\Delta E}$ to admit a UIE description, and will be used below as our first indicator.
For the convenience of discussion, we use $\Delta E^{(n)}_{\mathrm F}$ to denote the energy scale for which Eq.~\eqref{eq-Deltak-DE} is satisfied for all free cumulants up to order $n$, and one generally has
\begin{equation}
    \Delta E_U \le \Delta E_{\mathrm F}
    \equiv \Delta E^{(n\rightarrow\infty)}_{\mathrm F}
    \le \cdots
    \le \Delta E^{(2)}_{\mathrm F}
    \le \Delta E^{(1)}_{\mathrm F}.
\end{equation}

 As a complementary indicator, we also consider the relation introduced in Ref.~\cite{GETH-Foini19}. For real matrix elements, as is the case here, if ${\cal O}_{\Delta E_U}$ can be regarded as a sample from the UIE, then the probability distributions of its diagonal and off-diagonal elements satisfy
\begin{equation}\label{eq-uim-id2}
    P\left(\frac{{\cal O}_{\alpha\alpha}-{\cal O}_{\beta\beta}}{2}\right)=P({\cal O}_{\alpha\beta}) .
\end{equation}}

{We start with the indicator given in Eq.~\eqref{eq-Deltak-DE}.} Throughout the numerical simulations, we fix the center of the microcanonical energy window used in Eq.~\eqref{eq-PDE} at $E_0 = 0.0$.
Using standard sparse-matrix diagonalization, we obtain $2000$ eigenstates ($1000$ for $j\ge 4000$) and project the observables ${\cal A}$ and ${\cal B}$ onto the corresponding energy window. 
The free cumulants $\Delta_k$ are then calculated via Eq.~\eqref{eq-fc}.

The results for $\Delta_k$  as a function of energy window width $\delta E$ are shown in Figs.~\ref{Fig-FC},~\ref{Fig-FCB},~\ref{Fig-FC-Int} and \ref{Fig-FCB-Int}, where we {consider} observables ${\cal A}$ and ${\cal B}$ in {both} chaotic and integrable cases.
{The free cumulants exhibit markedly different behavior in the chaotic and (near-)integrable cases: the scaling
$\Delta_k \propto \delta E^{k-1}$
is observed only in the chaotic case, while clear deviations are found in the (near-)integrable case.}

\begin{figure}[t]
	\includegraphics[width=1.0\linewidth]{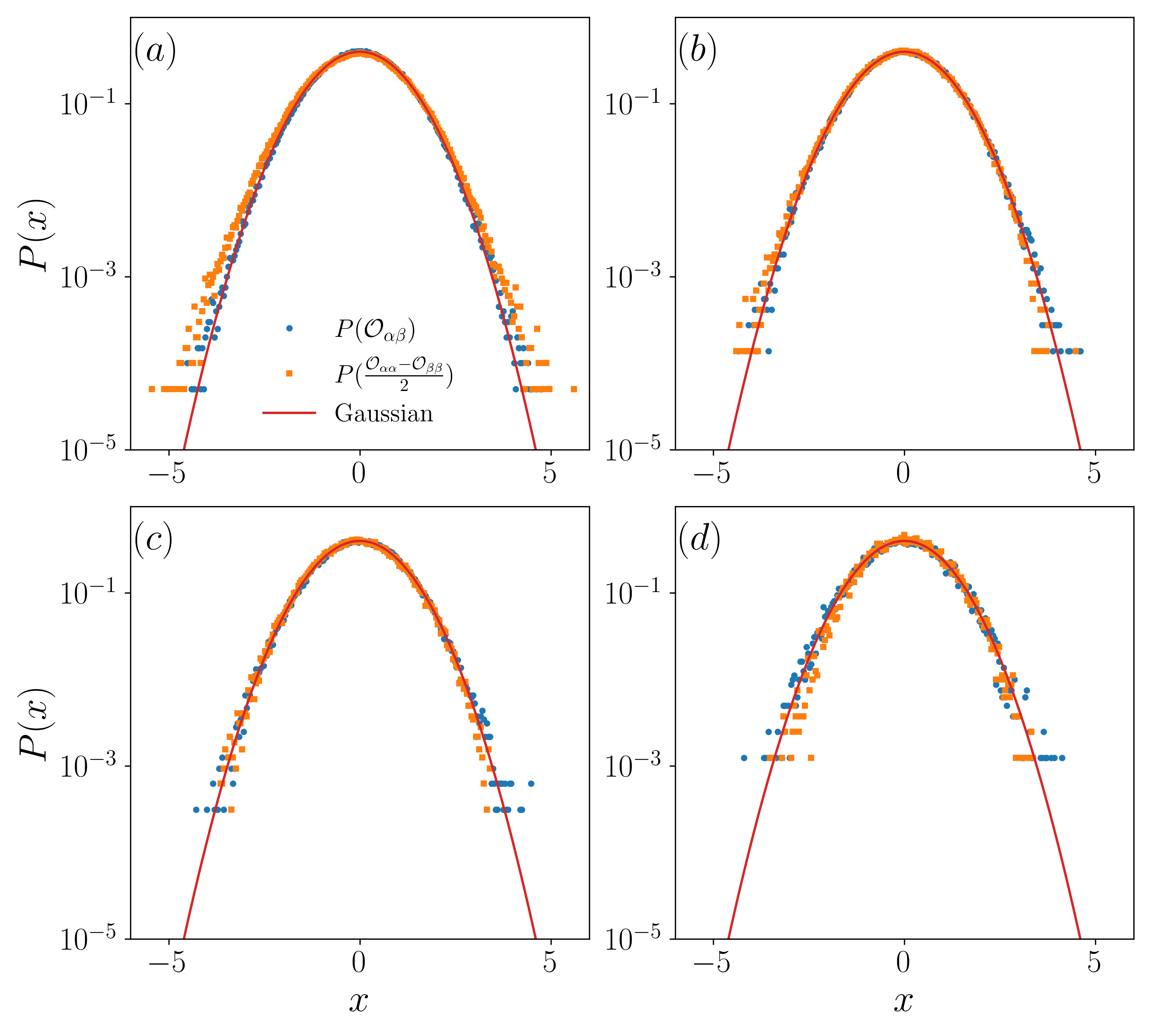}
 \caption{Indicator {for UIE description} of microcanonically truncated operators: $P({\cal O}_{\alpha\beta})$ versus $P(\frac{{\cal O}_{\alpha\alpha}-{\cal O}_{\beta\beta}}{2})$ for the chaotic case $\lambda = 0.0$ for matrix elements of ${\cal O}_{\delta E}$ for: (a) $\delta E = 1.2 \times 10^{-4}$; (b) $\delta E = 0.72 \times 10^{-4}$; (c) $\delta E = 0.48 \times 10^{-4}$ and (d) $\delta E = 0.24 \times 10^{-4}$.  
 Results are shown for operator ${\cal O} = {\cal A}$ and $j=4000$. The center of energy window is fixed at $E_0 = 0$. Red solid line indicates the Gaussian distribution.
 }\label{Fig-Dis-Amn}
\end{figure}

\begin{figure}[t]
	\includegraphics[width=1.0\linewidth]{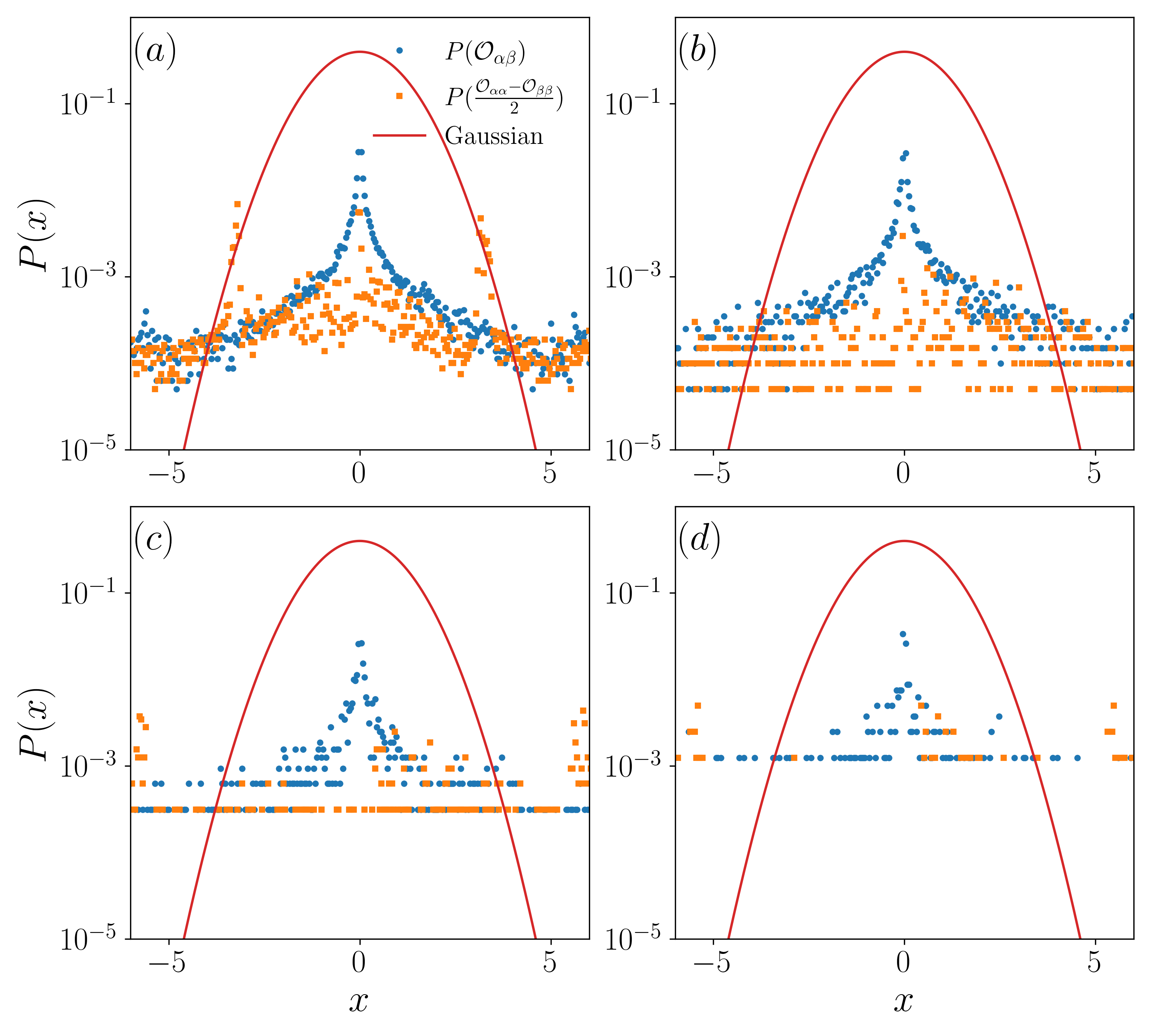}
 \caption{Indicator {for UIE description} of microcanonically truncated operators: $P({\cal O}_{\alpha\beta})$ versus $P(\frac{{\cal O}_{\alpha\alpha}-{\cal O}_{\beta\beta}}{2})$ for (near-)integrable case $\lambda = 0.75$ for matrix elements of ${\cal O}_{\delta E}$ for: (a) $\delta E = 4 \times 10^{-4}$; (b) $\delta E = 2 \times 10^{-4}$; (c) $\delta E = 8 \times 10^{-5}$ and (d) $\delta E = 4 \times 10^{-5}$. 
 Results are shown for operator ${\cal O} = {\cal A}$ and $j=3000$. The center of energy window is fixed at $E_0 = 0$. Red solid line indicates the Gaussian distribution.
 }\label{Fig-Dis-Amn-Int}
\end{figure}

If the energy window width $\delta E$ is further {reduced},
deviations from the {scaling relation} in Eq.~\eqref{eq-Deltak-DE}, $\Delta_k \propto \delta E^{k-1}$, become noticeable even in the chaotic case, when $\delta E$ is extremely small. However, as shown in Figs.~\ref{Fig-FC} and ~\ref{Fig-FCB}, the deviations
appear at smaller $\delta E / \hbar_{\text{eff}}$ for larger system size (larger $j$). 
A plausible explanation for these deviations is that Eq.~\eqref{eq-Deltak-DE} is derived under the condition that the number of states within the energy window $d_{\delta E}$ is sufficiently large. 
{Even if ${\cal O}_{\Delta E}$ admits a UIE description, deviations from the scaling predicted by Eq.~\eqref{eq-Deltak-DE} can appear when $d_{\delta E}$ is small.}
Further details are provided in Appendix \ref{app-2}, where we compute the averaged free cumulants in the UIE $\widetilde{\cal O}_{\Delta E} = U^\dagger {\cal O}_{\Delta E} U$, with $U$ denoting a Haar-random unitary operator. Similar deviations also appear at very small energy windows. {This suggests that the deviations from Eq.~\eqref{eq-Deltak-DE} observed for extremely small energy windows in the chaotic case are due to finite-size effects, which are expected to vanish as $j \to \infty$.}

{
Another important question concerns how the scaling energy scale $\Delta E_{\mathrm F}$, i.e., the energy scale up to which the relation in Eq.~\eqref{eq-Deltak-DE} holds, depends on $\hbar_{\mathrm{eff}}$.
In Figs.~\ref{Fig-FC} and \ref{Fig-FCB}, an approximate data collapse of $\Delta_k$ is observed when plotted against $\delta E / \hbar_{\mathrm{eff}}$ over an intermediate range of energy windows for free cumulants up to $k=8$.
However, the collapse becomes less pronounced as the boundary of the scaling regime is approached, making it difficult to determine the dependence of $\Delta E_{\mathrm F}$ on $\hbar_{\mathrm{eff}}$.
Nevertheless, compared to the results in Sec.~\ref{sec-eth-old}, it is evident that $\Delta E_{\mathrm F}$, and hence also $\Delta E_U$, is much smaller than the energy scale over which the conventional ETH remains valid.
}

As a further check, we consider the second indicator introduced in Eq.~\eqref{eq-uim-id2} and compare the distributions
$P\!\left(\frac{{\cal O}_{\alpha\alpha}-{\cal O}_{\beta\beta}}{2}\right)$
and
$P({\cal O}_{\alpha\beta})$,
where ${\cal O}_{\alpha\beta}$ are chosen from a microcanonical energy window of width $\delta E$.
Good agreement between these two distributions can be observed for $\delta E \le 0.48\times 10^{-4}$ in the chaotic case (Fig.~\ref{Fig-Dis-Amn}). In contrast, the opposite behavior is evident in the (near-)integrable case with $\lambda = 0.75$ in Fig.~\ref{Fig-Dis-Amn-Int}.

{In summary, our numerical results provide evidence consistent with the conjectured RMT universality picture for microcanonically truncated operators in quantum few-body systems with chaotic classical limits.}

\begin{figure*}
    \centering
    \includegraphics[width=0.9\linewidth]{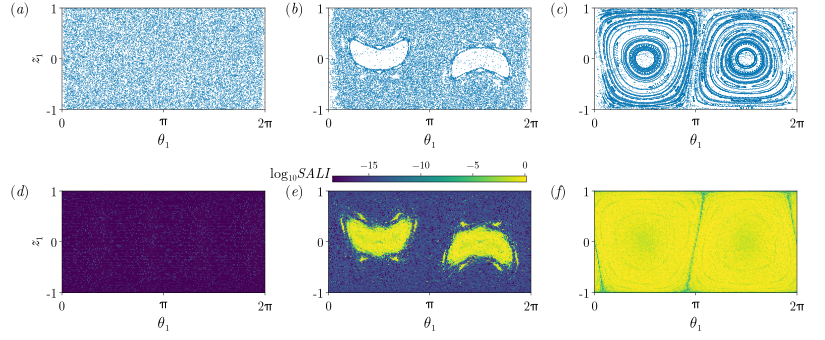}
    \caption{Comparison between the Poincar\'e sections in panels (a)-(c) and the SALI plot on the section in panels (d)-(f), for $\lambda=0,0.5,0.75$ from left to right, where at $\lambda=0.5$ the system exhibits mixed-type dynamics. The Poincar\'e sections are taken within the energy shell \( E = 0 \) at $\theta_2=\pi/2$. Panels (a) and (b) are generated by a single chaotic orbit evolved up to $t=10^5$, while panel (c) is generated by 100 random orbits up to $t=1500$. The SALI values in lower panels are computed up to \( t = 180 \).}
    \label{fig:pss-sali}
\end{figure*}

\section{Conclusions and Discussions}\label{sec-conclusion}
In this paper, using the Feingold--Peres model as an example, we have investigated the statistical properties of observables in quantum few-body systems with chaotic classical limits. Our main results are as follows:
(1) We proposed an ETH ansatz for quantum few-body systems with chaotic classical limits and verified it numerically.
(2) Going beyond the conventional ETH framework, we provided numerical evidence {consistent with the conjectured picture that truncated operators admit a UIE description within a sufficiently small energy window.}

{It is important to note that neither of the indicators employed in this work (Eqs.~\eqref{eq-Deltak-DE} and \eqref{eq-uim-id2}) provides a sufficient condition for establishing a UIE description of the truncated operator. Therefore, our results should not be regarded as a proof of such a description. A full proof remains a challenging problem and lies beyond the scope of the present work, representing an interesting direction for future investigation.}

{Other interesting directions include studying the full ETH in the present model, especially the structure of multi-point correlations, and investigating whether truncated operators in Floquet systems with chaotic classical limits, such as the kicked rotor and the kicked top, admit a UIE description within a certain quasi-energy window.}

\section*{Acknowledgements }
We thank M. Robnik and W.-g. Wang for fruitful discussions.
This work has been funded by the Deutsche
Forschungsgemeinschaft (DFG), under Grant No. 531128043, as well as under Grant
No.\ 397107022, No.\ 397067869, and No.\ 397082825 within the DFG Research
Unit FOR 2692, under Grant No.\ 355031190. H. Yan acknowledges the support of the Slovenian Research and Innovation
Agency (ARIS) under the grants J1-4387 and P1-0306.

\section*{Data Availability}
The data that support the findings of this article are openly
available \cite{wang_2025_16578978}.

\appendix
\section{Semiclassical density of states}
\label{app-0}
The Thomas-Fermi expression for the semiclassical density of states requires a phase-space integral to evaluate the classical phase-space volume, which is rarely available in closed form for general systems. For the coupled-spin system considered in this work, however, this integral $\rho_{\text{cl}}(E)=\frac{1}{(2\pi\hbar_\text{eff})^2}\int d\mathbf{z}d\mathbf{\theta}\ \delta(E - \mathcal{H}(\mathbf{z},\mathbf{\theta}))$ can be simplified, since 
\begin{align}
  I=\int& d\mathbf{z}d\mathbf{\theta}\ \delta(E - \mathcal{H}(\mathbf{z},\mathbf{\theta}))=\int d\mathbf{z}  d\mathbf{\theta}\delta(\eta-\xi\cos\theta_1\cos\theta_2)\nonumber\\
  &=\int d\mathbf{z} d\theta_1\sum_i 1/{|f'(\theta_2^{(i)})|},
\end{align}
where $f(\theta_2)=\eta-\xi\cos\theta_1\cos\theta_2$, and $\xi,\eta$ are given in Eq.~\eqref{eq-integ-parameters}, $\theta_2^{(i)}$ are the roots satisfying $\cos\theta_2=\eta/(\xi\cos\theta_1)$ which exist if and only if $|\eta/(\xi\cos\theta_1)|\le 1$. Since $f'(\theta_2)=\xi\cos\theta_1\sin\theta_2$, one obtains
\begin{align}
    I&=2\int d\mathbf{z}\int d\theta_1 \frac{\Theta(\xi^2\cos^2\theta_1-\eta^2)}{\sqrt{\xi^2\cos^2\theta_1-\eta^2}} \nonumber\\
    &=\int d\mathbf{z} \frac{8}{\xi}\int^{\pi/2}_{0}d\theta \frac{\Theta(\cos^2\theta-\eta^2/\xi^2)}{\sqrt{\cos^2\theta-\eta^2/\xi^2}} \nonumber\\
    &= \int d\mathbf{z} \frac{8}{\xi}\Theta(1-\eta^2/\xi^2)\int^{\arcsin a}_{0}d\theta \frac{1}{\sqrt{a^2-\sin^2\theta}}\nonumber \\
    &=\int d\mathbf{z} \frac{8}{\xi}K(1-\frac{\eta^2}{\xi^2})\Theta(1-\frac{\eta^2}{\xi^2}),
\end{align}
where $a^2=1-\eta^2/\xi^2$, $K(\cdot)$ denotes the complete elliptic integral of the first kind, and $\Theta(\cdot)$ denotes the Heaviside step function. We thus arrive at the semiclassical density of states given in Eq.~\eqref{eq-rho-cl}.

\begin{figure}
    \centering
    \includegraphics[width=1\linewidth]{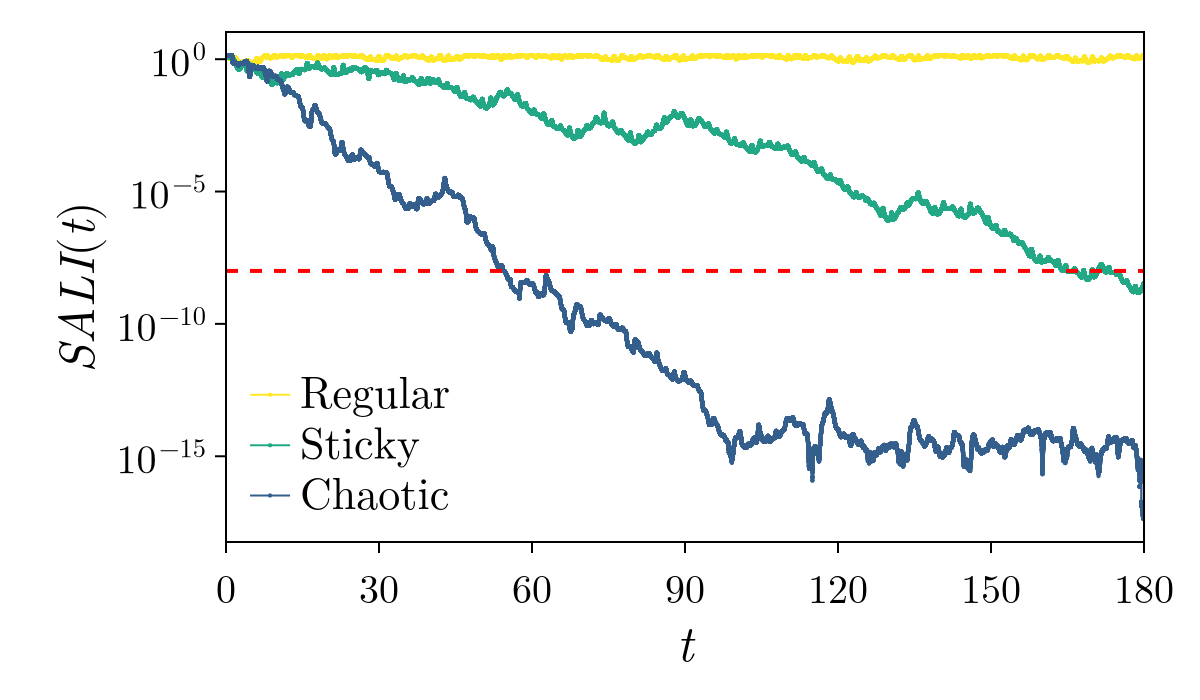}
    \caption{Evolution of the SALI over time \( t \) for three typical initial conditions selected from Fig.~\ref{fig:pss-sali} (e): a regular orbit, a sticky orbit at the boundary between chaotic and regular regions, and a chaotic orbit.}
    \label{fig:sali-detail}
\end{figure}

\section{Details of SALI in the system}
\label{app-1}
In the main text, using SALI, we show in Fig. \ref{Fig-DF} the presence of classical chaos and ergodicity across the energy spectrum. Here, we provide additional details and further evidence supporting this method. First, we define the characteristic function \(\chi_c\), which is 1 in chaotic regions and 0 otherwise.  Figure \ref{fig:pss-sali} compares Poincar\'e sections (for \(E = 0\) at \(\theta_2 = \pi/2\)) with SALI plots for \(\lambda = 0\) (fully chaotic), 0.5 (mixed-type), and 0.75 ((near-)integrable). The results show that SALI effectively distinguishes between regular and chaotic regions, including sticky regions at their boundaries, as further illustrated in Fig. \ref{fig:sali-detail}. We define \(\chi_c = 1\) for \(\text{SALI} \leq 10^{-8}\) at \(t = 180\), and 0 otherwise. For the numerical evaluation of $\mu_c$ defined in Eq. \eqref{eq-deltac}, we use the approximation:
 \begin{align}\label{eq-approx}
    \mu_c &= \frac{\int d\mathbf{z}d\mathbf{\theta} \, \chi_c(\mathbf{z},\mathbf{\theta}) \delta(E - \mathcal{H}(\mathbf{z},\mathbf{\theta}))}{\int d\mathbf{z}d\mathbf{\theta} \, \delta(E - \mathcal{H}(\mathbf{z},\mathbf{\theta}))}\nonumber\\
    &\approx 
    \frac{
\displaystyle
\sum_{ijk}\sum_r
w_{ijk}^{(r)}
\chi_c\!\left(
\phi_1^i,\phi_2^j,\theta_1^k,\theta_{2,ijk}^{(r)}
\right)
}{
\displaystyle
\sum_{ijk}\sum_r w_{ijk}^{(r)}
},
\end{align}
where $w_{ijk}^{(r)}
=\sin\phi_1^i\sin\phi_2^j/\left|
\partial_{\theta_2}\mathcal{H}
\right|_{\theta_2=\theta_{2,ijk}^{(r)}}
$, and for each sampled point $(\phi_1^i,\phi_2^j,\theta_1^k)$, $\{\theta_{2,ijk}^{(r)}\}$ denote solutions for $E=\mathcal{H}(\mathbf{z},\theta)$ with $z_1=\cos\phi_1^i,z_2=\cos\phi_2^j$. The Monte Carlo method is an effective way to compute this approximation, as shown in the main text.

\section{Detailed discussion of Eq.~\eqref{eq-ETH-new1}}\label{sec-appendixb}
To illustrate the scaling of $f(\overline{E},\omega)$ with $\hbar_{\text{eff}}$ indicated in Eq.~\eqref{eq-ETH-new}, let us consider the auto-correlation function of ${\cal O}$ 
\begin{equation}
   C(t)=\frac{1}{{\cal D}}\text{Tr}[{\cal O}(t){\cal O}]=\frac{1}{{\cal D}}\sum_{\alpha\beta}|{\cal O}_{\alpha\beta}|^{2}e^{-i(E_{\beta}-E_{\alpha})t/\hbar_{\text{eff}}} . 
\end{equation}
In the semiclassical limit $\hbar_{\text{eff}} \rightarrow 0$, 
\begin{align}
    C(t)\simeq C_{\text{cl}}(t)\equiv\frac{1}{(4\pi)^{2}}\int d\boldsymbol{z}d\boldsymbol{\theta}{\cal O}^{\text{cl}}(\boldsymbol{z}(t),\boldsymbol{\theta}(t)) \nonumber\\
    \quad \times\ {\cal O}^{\text{cl}}(\boldsymbol{z}(0),\boldsymbol{\theta}(0)).
\end{align}
Particularly, we are interested in 
\begin{equation}
    \widetilde{C}(t)=C(t)-C(\infty) = \frac{1}{{\cal D}}\sum_{\alpha\neq\beta}|{\cal O}_{\alpha\beta}|^{2}e^{-i(E_{\beta}-E_{\alpha})t/\hbar_{\text{eff}}} ,
\end{equation}
and its Fourier transform
{\begin{align}
\text{FT}\left[\widetilde{C}(t)\right] & =\frac{1}{{\cal D}}\sum_{\alpha\neq\beta}|{\cal O}_{\alpha\beta}|^{2}\int dt e^{-i(E_{\beta}-E_{\alpha})t/\hbar_{\text{eff}}-i\nu t} \nonumber \\
    & = \frac{2\pi}{{\cal D}}\sum_{\alpha \neq \beta}|{\cal O}_{\alpha\beta}|^{2}\delta(\nu-(E_{\alpha}-E_{\beta})/\hbar_{\text{eff}}) .
\end{align}}
Inserting Eq.~\eqref{eq-ETH-new0}, one obtains
\begin{align}
&\text{FT}\left[\widetilde{C}(t)\right]\simeq\frac{2\pi}{{\cal D}}\int d\overline{E}d\omega\frac{\rho(\overline{E}+\frac{\omega}{2})\rho(\overline{E}-\frac{\omega}{2})}{\rho(\overline{E})}|f(\overline{E},\omega)|^{2} \nonumber \\
 &    \qquad \qquad    \qquad \qquad    \qquad \qquad    \qquad \qquad     \times\ \delta(\nu-\omega/\hbar_{\text{eff}}) \nonumber \\
    \simeq & \frac{2\pi\hbar_{\text{eff}}}{{\cal D}}\int d\overline{E}\rho(\overline{E})|f(\overline{E},\hbar_{\text{eff}}\nu)|^{2} = 2\pi \hbar_{\text{eff}} \overline{|f(\overline{E},\hbar_{\text{eff}}\nu)|^{2}},
\end{align}
where
\begin{equation}
    \overline{|f(\overline{E},\hbar_{\text{eff}}\nu)|^{2}}\equiv\frac{1}{{\cal D}}\int d\overline{E}\rho(\overline{E})|f(\overline{E},\hbar_{\text{eff}}\nu)|^{2} .
\end{equation}
In the semiclassical limit $\hbar_{\text{eff}} \rightarrow 0$,
\begin{equation}
\text{FT}\left[\widetilde{C}(t)\right]\simeq\text{FT}\left[\widetilde{C}_{\text{cl}}(t)\right]\equiv F_{\text{cl}}(\nu) ,
\end{equation}
which leads to
\begin{equation}\label{eq-f2}
    \overline{|f(\overline{E},\hbar_{\text{eff}}\nu)|^{2}}\simeq\frac{1}{2\pi\hbar_{\text{eff}}}F_{\text{cl}}(\nu) .
\end{equation}
Letting $\omega = \hbar_{\text{eff}} \nu$, Eq.~\eqref{eq-f2} becomes
\begin{equation}
    \overline{|f(\overline{E},\omega)|^{2}}\simeq\frac{1}{2\pi\hbar_{\text{eff}}}F_{\text{cl}}(\omega/\hbar_{\text{eff}}),
\end{equation}
which supports the scaling of $f(\overline{E}, \omega)$ given in Eq.~\eqref{eq-ETH-new}.


\begin{figure}
\centering	\includegraphics[width=1.0\linewidth]{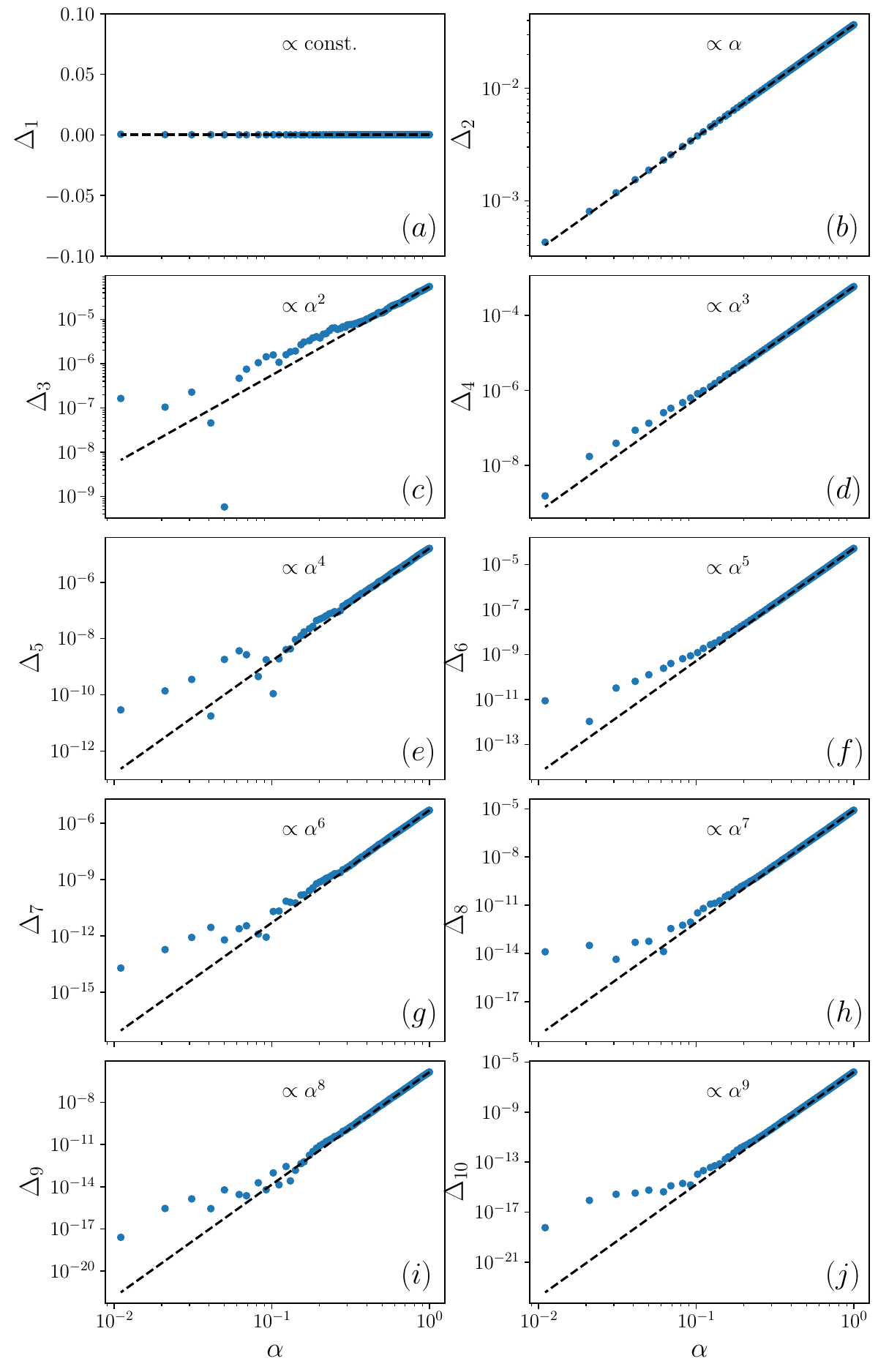}
 \caption{Free cumulants $\Delta_k$ versus relative dimension $\alpha = d/d_U$ in the UIE defined in Eq.~\eqref{eq-Au}.
The dashed line indicates the predicted scaling $\Delta_k \propto \alpha^{k-1}$.
We take the  average over $100$ samples in the ensemble by considering different realizations of $U$.}\label{Fig-FCR}
\end{figure}

\section{Derivation of Eq.~\eqref{eq-fca}}\label{apx-fc}
In this appendix, we 
give an alternative derivation of Eq.~\eqref{eq-fca} different from the one given in Ref.~\cite{wang2023emergence}.

We begin by recalling the definition of UIE introduced in Eq.~\eqref{eq-uie}
\begin{equation}
{\cal O}_{U}=U{\cal O}^{\star}U^{\dagger},
\end{equation}
with $U$ being a Haar-random unitary operator.
We consider $\cal O$ as a typical sample of ${\cal O}_U$, with a Hilbert space dimension $D$.
For sufficiently large $D$, in a chosen basis ${|i\rangle}$, the $k$-th free cumulant of 
$\cal O$ can be written as
\begin{equation}\label{eq-Deltak-Oab}
\Delta_{k}=\frac{1}{D}\sum_{i_{1}\neq i_{2}\neq\cdots\neq i_{k}}{\cal O}_{i_{1}i_{2}}{\cal O}_{i_{2}i_{3}}\cdots{\cal O}_{i_{k}i_{1}}\ ,
\end{equation}
where ${\cal O}_{ij}=\langle i|{\cal O}|j\rangle$.
In this case,  one may replace 
${\cal O}_{i_{1}i_{2}}{\cal O}_{i_{2}i_{3}}\cdots{\cal O}_{i_{k}i_{1}}$ in the summation of Eq.~\eqref{eq-Deltak-Oab} with its UIE average, 
$\overline{{\cal O}_{i_{1}i_{2}}{\cal O}_{i_{2}i_{3}}\cdots{\cal O}_{i_{k}i_{1}}}$. According to Ref. \cite{Maillard_2019_geth}, 
\begin{equation}\label{eq-O-uim}
\overline{{\cal O}_{i_{1}i_{2}}{\cal O}_{i_{2}i_{3}}\cdots{\cal O}_{i_{k}i_{1}}}=\frac{1}{D^{k-1}}\Delta_{k}=\text{const.}\ .   
\end{equation}
Now we truncate $\cal O$ to a $d$-dimensional subspace of the Hilbert space, denoted by $S_d$, by introducing the projection operator
\begin{equation}
P_{d}=\sum_{i\in S_{d}}|i\rangle\langle i| \, ,
\end{equation}
which yields
\begin{equation}\label{eq-Od}
{\cal O}_{d}=P_{d}{\cal O}P_{d}\ .
\end{equation}
For sufficiently large $d$, similar to Eq.~\eqref{eq-Deltak-Oab}, the $k$-th free cumulant of ${\cal O}_{d}$ reads
\begin{equation}\label{def-PBRM}
\Delta_{k}(d)=\frac{1}{d}\sum_{\substack{i_{1}\neq i_{2}\neq\cdots\neq i_{k}\\
i_{1}\in S_{d,}\ldots i_{k}\in S_{d}
}
}{\cal O}_{i_{1}i_{2}}{\cal O}_{i_{2}i_{3}}\cdots{\cal O}_{i_{k}i_{1}} .
\end{equation}
Using a similar argument and inserting Eq.~\eqref{eq-O-uim}, one obtains
\begin{equation}\label{eq-Od-uim}
    \Delta_{k}(d)\simeq\frac{1}{d}\sum_{\substack{i_{1}\neq i_{2}\neq\cdots\neq i_{k}\\
i_{1}\in S_{d,}\ldots i_{k}\in S_{d}
}
}\frac{\Delta_{k}}{D^{k-1}}\simeq \left(\frac{d}{D}\right)^{k-1}\Delta_{k}\ ,
\end{equation}
which coincides with Eq.~\eqref{eq-fca}.
{Before ending this section, we emphasize that Eq.~\eqref{eq-Od-uim} should be regarded as a necessary condition for a UIE description, rather than a sufficient one.}

\section{Deviation from Eq.~\eqref{eq-Deltak-DE} in extremely small energy window in the chaotic case} 
\label{app-2}
In this section, we provide further discussion on the reason for the deviations from Eq.~\eqref{eq-Deltak-DE} in extremely small energy window in the chaotic case, observed in Fig.~\ref{Fig-FC} and Fig.~\ref{Fig-FCB}. To this end, we construct a unitarily invariant ensemble, 
$\widetilde{\cal O}_{\Delta E} = U^\dagger {\cal O}_{\Delta E} U$, where $U$ is a $d_{\Delta E} \times d_{\Delta E}$ Haar random unitary operator. Practically, we choose the ``seed" operator ${\cal O}_{\Delta E} = {\cal A}_{\Delta E} = P_{\Delta E} {\cal A} P_{\Delta E}$ which is considered in Fig.~\ref{Fig-FC}, where the energy window consists of $d_\text{U}=d_{\Delta E} = 1000$ eigenstates with energy $E\approx 0$ and $j = 5000$. 
{The UIE is thus constructed as
\begin{equation}\label{eq-Au}
\widetilde{{\cal A}}_\text{U}=U^{\dagger}{\cal A}_{\Delta E}U\ .
\end{equation}}
We then project it onto a smaller subspace with dimension $d$, which yields ${\cal A}_d = P_d \widetilde{{\cal A}}_{\text{U}} P_d$. 
Since dimension is more essential in the discussion of UIE, we denote the projection operator and the truncated operator using $d$ instead of the energy scale. 
If $d$ is sufficiently large, the ensemble average of free cumulants of ${\cal A}_d$ follows Eq.~\eqref{eq-fca} 
\begin{equation}\label{eq-uim-d}
     {\Delta_{k}^{d}}=\alpha^{k-1}{\Delta_{k}^{U}},\ \alpha = d/d_\text{U} . 
\end{equation}
{This} is confirmed by the numerical results in Fig.~\ref{Fig-FCR} for $\alpha\gtrsim 0.5$ ($d\gtrsim 500$).
However, when $\alpha$ (as well as $d$) is sufficiently small, i.e., $\alpha\lesssim0.2$ ($d\lesssim200$), deviations from Eq.~\eqref{eq-uim-d} become visible, similar to the deviations found in Figs.~\ref{Fig-FC} and ~\ref{Fig-FCB}. 
{This suggests that the deviations from Eq.~\eqref{eq-Deltak-DE} observed in the main text at extremely small energy windows are likely due to the small dimension of the Hilbert subspace associated with the corresponding energy window, rather than indicating a breakdown of the UIE description in this case.}

%

%

\end{document}